\documentclass[twocolumn,aps,pre,showpacs,preprintnumbers]{revtex4}
\RequirePackage{ifpdf} 
\usepackage[dvipdf]{graphicx}
\usepackage{epstopdf}
\usepackage{epsfig} 
\usepackage{subfigure} 
\usepackage{hyperref}
\usepackage{graphics}
\usepackage{color}      
\usepackage{dcolumn}
\usepackage{amsmath,mathrsfs}

\begin{document}
\title{First passage time and stochastic resonance of  excitable systems}

\author{Solomon Fekade Duki}
\affiliation {National Center for Biotechnology Information,
National Library of Medicine and National Institute of Health,
8600 Rockville Pike, Bethesda MD, 20894 USA}
\author{Mesfin Asfaw Taye}
\affiliation{Department of Physics, California
State University  Dominguez Hills, California, USA }

\begin{abstract} 

We study noise induced  thermally activated barrier crossing of a Brownian particle 
that hops in a periodic ratchet potential where the ratchet potential is coupled with 
a spatially uniform temperature. The viscous friction $\gamma$ is considered to decrease 
exponentially when the temperature $T$ of the medium increases  ($\gamma=Be^{-A T}$) 
as proposed originally by Reynolds \cite{am10}.  The results obtained in this work 
show that  the mean first passage time  of the particle  is considerably lower when 
the viscous friction is temperature dependent than that of the case where the viscous 
friction is temperature independent. Using  exact analytic solutions and via numerical 
simulations not only we  explore the dependence for the mean first passage time of  a 
single particle  but also we  study the dependence for the first arrival time of one 
particle out of many particles.  Our result exhibits that the first  arrival time 
decreases as the number of particles increases. We then explore the thermally activated 
barrier crossing rate of the system in the presence of time varying signal. In this case, 
the interplay between noise and  sinusoidal driving force in the bistable system may lead 
the system into stochastic resonance provided that  the random tracks are adjusted in an optimal
way to the recurring external force. The dependence of signal to noise ratio $SNR$  as well 
as the power ampliﬁcation ($\eta$) on model parameters is explored. $\eta$ as well as SNR 
depicts a pronounced peak at a particular noise strength $T$. The magnitude of $\eta$ is 
higher for temperature dependent $\gamma$ case. In the presence of $N$ particles,  $\eta$ 
is considerably amplified as $N$ steps up showing the the weak periodic signal plays a vital 
role in controlling the noise induced dynamics of excitable systems.
\end{abstract}
\pacs{Valid PACS appear here}
\maketitle
 
\section {Introduction}

Studying the mean first passage time (MFPT) of various physical problems is vital and has diverse 
applications in many disciplinary fields such as science and engineering. In most cases, the MFPT 
is  usually defined as the amount of time that a given particle takes to surmount  a certain 
threshold where the threshold can be specified as a certain boundary,  potential barrier and specified state.  Particularly if one considers a Brownian particle moving in a viscous medium, assisted 
by the thermal background kicks, the particle presumably crosses the potential barrier. The magnitude of its 
MFPT relies not only on the system parameters, such as the potential barrier height, but also  it
depends on the initial and boundary conditions.  Understanding of such noise induced  thermally 
activated barrier crossing problem  is vital  to get a better understanding of most biological 
problems  \cite{am1,am2,am3,am4,am5,am6,am7,am8}. In the past, considering 
temperature independent viscous friction, the dependence of the mean first passage time (equivalently
the escape rate) on model parameters has been explored for  various model systems, see for example the work \cite {am8,am9,am24}. 
However experiment shows that the viscous friction $\gamma$ is indeed temperature dependent and 
it  decreases as temperature increases. In this work we discuss the role of temperature on the 
viscous  friction as well as on the MFPT  by taking a viscous friction $\gamma$ that  decreases 
exponentially when the temperature $T$ of the medium increases  ($\gamma=Be^{-A T}$) as proposed 
originally by Reynolds \cite{am10}. It is shown that the MFPT is smaller in magnitude when 
$\gamma$ is temperature dependent than when it is temperature independent. This is plausible 
since the diffusion constant $D=T/\gamma\propto k_{B}T e^{AT}$  is valid when  the viscous friction considered to be 
temperature dependent showing that the effect of temperature on the particle mobility 
is twofold. First, it directly assists the particle to surmount the potential
barrier. In other words, the particle jumps the potential barrier at the expenses of the thermal 
kicks. Second, when temperature increases, the viscous friction gets attenuated and as a result 
the diffusibility of the particle increases.

The first passage time problem   has also been extensively studied in many excitable systems such as chemical reaction,  neural system
and cardiac system \cite {am111, am11,am12}. Particularly in cardiac system,  the intra-cellular calcium 
dynamics is responsible for a number of trigged arrhythmias \cite{am12}. As discussed in our previous work \cite{am12}, the abnormal calcium 
release at a single microdomain level   can be studied 
via master equation, where the corresponding Fokker-Planck equation can be written with an 
effective bistable potential. The MFPT for a single Brownian 
particle to cross the effective potential then corresponds to the time it takes for $n$ channels 
to open at a single microdomain level. Thus, although  in the present paper we consider a 
simplified ratchet potential, our study gives us a clue regarding the dynamics of 
calcium ions in the cardiac system. Moreover membrane depolarization occurs if the simulations 
happen on tissue level when $N$ microdomains interact. The First passage time for  one of 
these $N$ microdomains  to fire  for the     first  time can be found by calculating  the MFPT 
that  one particle takes out of $N$ particle  to cross the potential barrier.

Exposing  excitable systems to time varying periodic  forces may result in an intriguing 
dynamics where in this case the coordination of the noise with time varying force leads to 
the phenomenon of stochastic resonance (SR) \cite{am13,am14} provided that the noise induced 
hopping events synchronize with the signal. The phenomenon of stochastic resonance 
has obtained considerable interests because of its signiﬁcant practical applications in 
a wide range of fields. SR depicts that systems  enhance their performance as long as   
the thermal background noise  is synchronized with time varying  periodic signal. Since 
the innovative work of Benzi {\em et. al.} \cite{am13}, the idea of stochastic resonance has been 
broadened and implemented to many  model  systems \cite{am15,am16,am17,am18,am19,am20,am21,am22,am23}.  
Recently the occurrence  of stochastic resonance for a Brownian particle  as well as 
for extended system such as polymer has been reported by us \cite{am24,am25}.  Our analysis 
revealed that, due to the ﬂexibility that can enhance crossing rate and change in chain 
conformations at the barrier, the power ampliﬁcation exhibits an optimal value at optimal 
chain lengths  and elastic constants  as well as at optimal noise strengths. However 
most of these studies considered a viscous friction which is temperature independent. 
In this work, considering temperature dependent viscous friction, we study how the 
power ampliﬁcation behaves as one varies the model parameters. We first explore the 
stochastic resonance of a single particle and  we 
then study the SR for  many particle system by considering both 
temperature dependent and  independent viscous friction cases.

The aim of this paper is to explore the crossing rate and stochastic resonance of a 
single as well as many Brownian particles in a piecewise linear bistable potential by 
considering both temperature dependent and independent viscous friction cases.  
Although a generic model system is considered,  the present study   helps to 
understand the dynamics of excitable systems and it is also vital for basic understanding
of statistical physics. The MFPT at single particle level is 
extensively studied in the past see for example the work \cite{am9}. However, the role of temperature on viscosity as well as on MFPT 
has not been studied in detail and  this will be the subject of the present paper. Particularly, in the presence
of time varying signal,  we study  how the background temperature affects the viscosity as well as the signal to 
noise ratio $SNR$  and spectral density$ \eta$.  On the other hand, the first  passage time statistics at ensemble  ($N$ particles) level 
has been explored in many studies \cite{am11,am12}.  However, to best of our knowledge, the role of time-varying signal as well as 
the role of temperature on $SNR$ and $\eta$ has not been studied in detail at the ensemble level. In this work, via numerical simulations and using the exact analytic results, we study stochastic resonance 
of $N$ particles.

To give you a brief outline, in this work first we study the MFPT of a single particle
both for temperature dependent and independent viscous friction cases. The exact analytic 
results as well as the simulation results depict that the MFPT is considerably smaller 
when $\gamma$ is temperature dependent.  In both cases the escape rate increases as 
the noise strength increases and  decreases as the potential barrier increases.  We 
then  extend our study for $N$ particle systems. The First passage time for  one of 
the $N$ particles to fire for the first time $T_N$ can be found both analytically 
(at least in the high barrier limit) and via numerical simulation for a bistable system. 
It is found that $T_{N}$ is considerably  smaller when the viscous friction is 
temperature dependent. For both cases, $T_{N}$ decreases as the noise strength increases 
and as the potential barrier steps down.  In high barrier limit, $T_{N}=T_{s}/N$ 
where $T_{s}$ is the MFPT for a single particle. In general as the number of particles
$N$ increases, $T_{N}$ decreases.

We then study our  model system in the presence of time varying signal. In this case 
the interplay between noise and sinusoidal driving force in the bistable system may 
lead the system into stochastic resonance. Analytically and via numerical simulations, 
we study how the signal to noise ratio (SNR) and  power ampliﬁcation ($\eta$) behave 
as a function of the model parameters. $\eta$ as well as SNR depicts a pronounced peak
at particular noise strength $T$. The magnitude of $\eta$ is higher for temperature
dependent $\gamma$ case. In the presence of many particles $N$,  $\eta$ is considerably 
amplified as $N$ steps up, showing that the weak periodic signal plays a vital   
role in controlling the noise induced dynamics of excitable systems.

The rest of the paper is organized as follows. In section II, we present the model. 
In section III, by considering both temperature dependent and independent viscous 
friction cases, we explore the dependence for MFPT on model parameters for a
single as well as many particle systems. The role of  sinusoidal driving force on 
enhancing the mobility of the particle is studied in IV. Section V deals with 
Summary and Conclusion.

\section{The model}  
Let us consider a Brownian particle that walks in a piecewise linear potential with an 
external load $U(x)$, where the ratchet potential $U(x)$ is given by
\begin{equation}
  U(x)=\left\{
  \begin{array}{ll}
    U_{0}\left({~x\over L_{0}}+1\right),& \text{if}~~~ -L_0 \le x \le {0};\\ 
    U_{0}\left(-{x\over L_{0}}+1\right),& \text{if}~~~{0} \le x \le L_{0}. 
   \end{array}
   \right.
\end{equation}
Here $U_{0}$ and $2L_{0}$ denote  the barrier height and the width of the ratchet potential,
respectively. The potential exhibits its maximum value $U_{0}$ at  $x={0}$ and its minima 
at $x=-L_0$  and $x=L_{0}$. The ratchet potential is coupled with a uniform  temperature $T$
as shown in Fig. 1.  
\begin{figure} 
 \includegraphics[width=10cm]{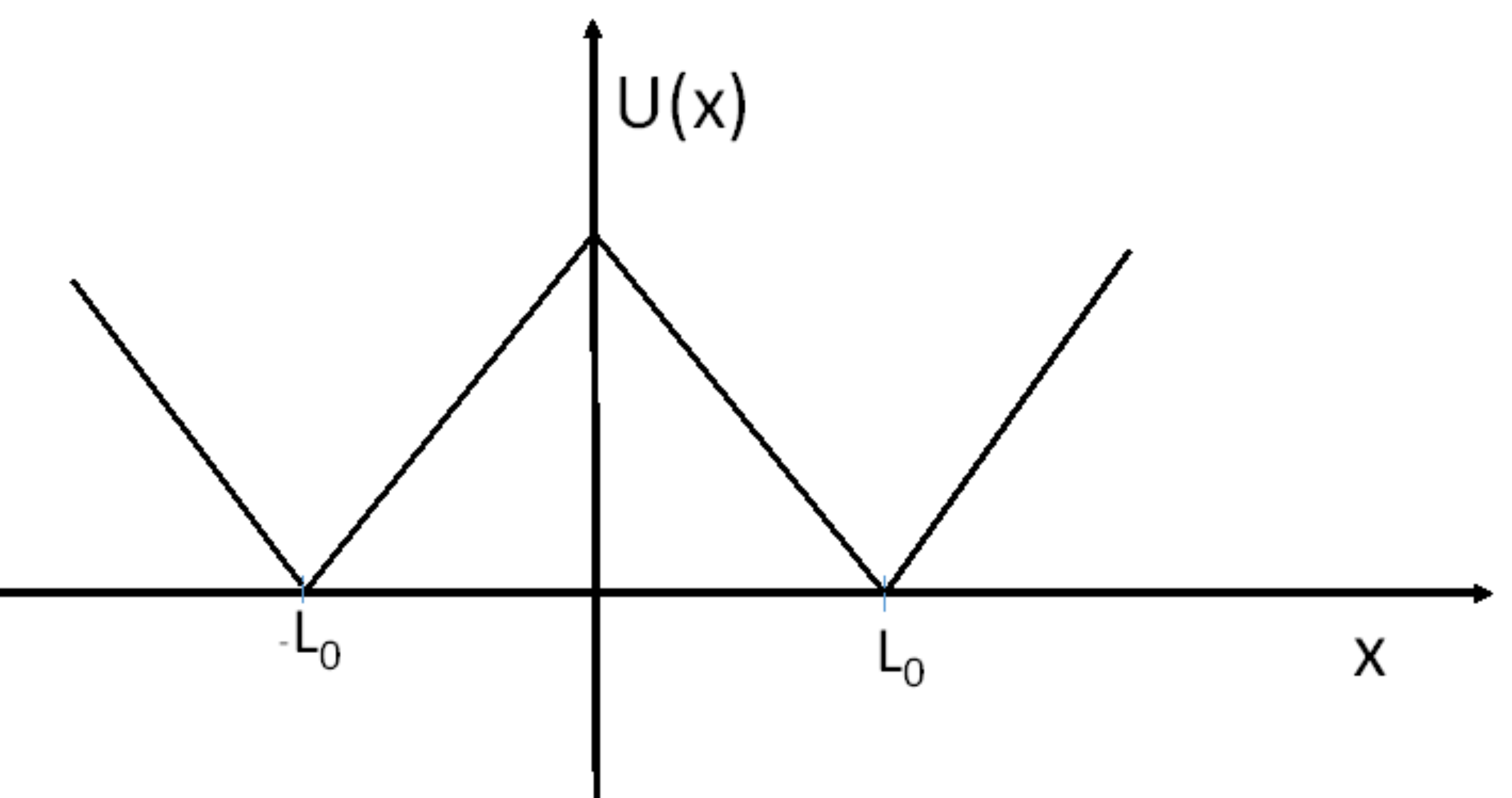}
\caption{ Schematic diagram for a Brownian particle in a piecewise linear  potential. 
Due to  the thermal background kicks, the particle ultimately surmounts the potential barrier.}
\end{figure}

For a Brownian particle that is arranged to undergo a random walk in a highly
viscous medium, the dynamics of the particle is governed  by  Langevin equation \cite{am1}. 
The general stochastic Langevin equation, which is derived in the  pioneering work 
of Petter H\"anggi \cite{am2}, can be written as
\begin{eqnarray}
\gamma{{dx}\over {dt}} = - \partial_x U -{(1-\epsilon)\gamma^{-1}}
{\partial}_x(\gamma T)+ \sqrt{2k_{B}\gamma T}\xi(t)
\end{eqnarray}
where $\gamma=\gamma(x)$ is the viscous friction, and $k_B$ is the Boltzmann's constant \cite{am3}. 
The  It\'o  and  Stratonovich interpretations correspond to the case where $\epsilon=1$  and $\epsilon=1/2$, 
respectively while the case  $\epsilon=0$ is   known as  the  H\"anggi   a post-point
or transform-form interpretation. At this point we want to stress that since we 
consider a uniform  temperature  profile,  the expressions for  thermodynamic 
quantities  do not  depend on the type of interpretation we use which implies 
the term ${(1-\epsilon)\over \gamma(x)}{\partial\over   \partial x}(\gamma(x)T(x))$ 
can be omitted. Here after we adapt the Langevin equation 
\begin{eqnarray}
\gamma{dx\over dt}&=& - {\partial}_x U(x) + \sqrt{2k_{B}\gamma(x) T}\xi(t).
\end{eqnarray}
The viscous friction has  an exponential temperature dependence 
 \begin{eqnarray}
 \gamma & =    Be^{-A T},& \text{if}~~~ -L_0 \le x \le {L_{0}}.
\end{eqnarray}
where $A$ and $B$ are constants. The random noise $\xi(t)$ is assumed to be Gaussian white noise 
satisfying the relations $\left\langle  \xi(t) \right\rangle =0$ and $\left\langle \xi(t)  \xi(t')
\right\rangle=\delta(t-t')$  where $k_{B}$  and $B$ are considered to be unity.

In the high friction limit, the dynamics of the Brownian particle is governed by 
\begin{equation}
{\partial P(x,t)\over \partial t}={\partial\over  \partial x}
\left[{U'(x)\over \gamma}P(x,t)+{\partial \over \partial x}\left({T\over \gamma}P(x,t)\right)\right]
\end{equation}
where $P(x,t)$ is the probability density of finding the particle at position $x$ and  time $t$. Here
$U'(x)={d\over dx}U$.  At stationary state 
$J(x) = -\left[{U'(x)\over \gamma}P_{s}(x) + {\partial \over \partial x}
\left({T\over {\gamma }}P_{s}(x)\right)\right]$.

The diffusion constant $D={{k_BT}\over {\gamma}} = k_BT e^{AT}$  is valid when  viscous friction to 
be temperature dependent showing that the effect of temperature on the particles' mobility 
is twofold. First, it directly assists the particles to surmount the potential
barrier; {\em i. e.} particles jump the potential barrier at the expenses of the thermal 
kicks. Second, when temperature increases, the viscous friction gets attenuated and as a 
result the diffusibility of the particle increases. Various experimental studies also showed
that the viscosity  of the  medium  tends to decrease as the  temperature of the medium 
increases. This is because increasing the temperature steps up the speed of the 
molecules, and this in turn creates   a reduction in the interaction time between
neighboring  molecules.  As a result, the intermolecular force between the molecules 
decreases and hence  the magnitude of the viscous friction decreases. Next we look at 
the dependence of the first passage time on the model parameters. 

Hereafter, all the figures are plotted using the following dimensionless parameters:
temperature ${\bar T}(x)=T (x)/T_{c}$, barrier height ${\bar U_{0}}=U_{0}/T_{c}$ and  
length ${\bar x}=x/L_{0}$. Moreover, all equations will be expressed in terms of the 
dimensionless parameters and for brevity we drop  all the bars hereafter.

\section{The Mean first passage time  of a single 
and many non-interacting particles }

\subsection{Mean first passage time for a single Brownian particle}

We consider a single Brownian particle  which is initially placed on the local minimum 
of a linear bistable potential as shown in Fig. 1. Due to the thermal background kicks, the 
particle  presumably crosses the potential barrier. The magnitude of the crossing rate 
of the particle strictly relies on the barrier height and noise strength as well 
as on the length of the ratchet potential. 

The mean first passage time $T_{s}$  for Brownian particle that walks on the ratchet potential can be found via 
\begin{eqnarray}
T_s=\int_{-L_0}^{L_0}dx e^{U(x) \over T}\int_{-L_0}^{x}dz {e^{-U(z) \over T}\over h}
\end{eqnarray}
where $h = {T \over \gamma}$ \cite{am26}. If one imposes  a reflecting boundary condition at $x=-L_{0}$
and absorbing boundary condition at $x=L_0$, Eq. (6) converges to 
\begin{eqnarray}
T_s = T_1 + T_2
\end{eqnarray}
where
\begin{eqnarray}
T_1&=&\int_{-L_0}^0dx e^{U^1(x)\over T}\int_{-L_0}^{x}dz {e^{-U^1(z)\over T}\over h}\nonumber \\
&=&{L_{0}^2 [T (-1 + e^{U_0\over T}) - U_0]\over U_{0}^2}e^{-AT}
\end{eqnarray}
and
\begin{eqnarray}
T_{2}&=&
\int_{0}^{L_{0}}dx e^{U^2(x)\over T}\int_{-L_0}^{0}dz {e^{-U^1(x)\over T}\over h}+\nonumber \\
 &&\int_{0}^{L_{0}}dx e^{U^2(x)\over T}\int_{0}^{L_0}dz {e^{-U^2(z)\over T}\over h}\nonumber \\
&=&{{L_{0}^2 [U_{0} + T (-3 + 3 \cosh[{U_0\over T}] - \sinh[{U_0\over T}])]}\over {U_{0}^2}}e^{-AT}
\end{eqnarray}
where $U^1(x)=U_{0}\left({x\over L_{0}}+1\right)$ and $U^2(x)=U_{0}\left({{-x}\over L_{0}}+1\right)$.
After some algebra we find 
\begin{equation}
T_s={4TL_{0}^2e^{-AT}\left(-1+\cosh[{U_0\over T}]\right)\over U_{0}^2}.
\end{equation}
Equation (10) is an exact analytic expression and its validity is justified using  numerical 
simulations. In  high barrier limit $U_0 \to \infty$, $T_s$ approaches
\begin{equation}
T_s={2TL_{0}^2e^{-AT}e^{{U_0\over T}}\over U_{0}^2}.
\end{equation}
For temperature independent viscous friction ($A=0$), one retrieves 
\begin{equation}
T_s={4TL_{0}^2\left(-1+\cosh[{U_0\over T}]\right)\over U_{0}^2}.
\end{equation}
In  high barrier limit $U_0 \to \infty$, $T_s$  approaches
\begin{equation}
T_s={2TL_{0}^2e^{{U_0\over T}}\over U_{0}^2}.
\end{equation}

\begin{figure}[ht]
\centering
{
    \includegraphics[width=6cm]{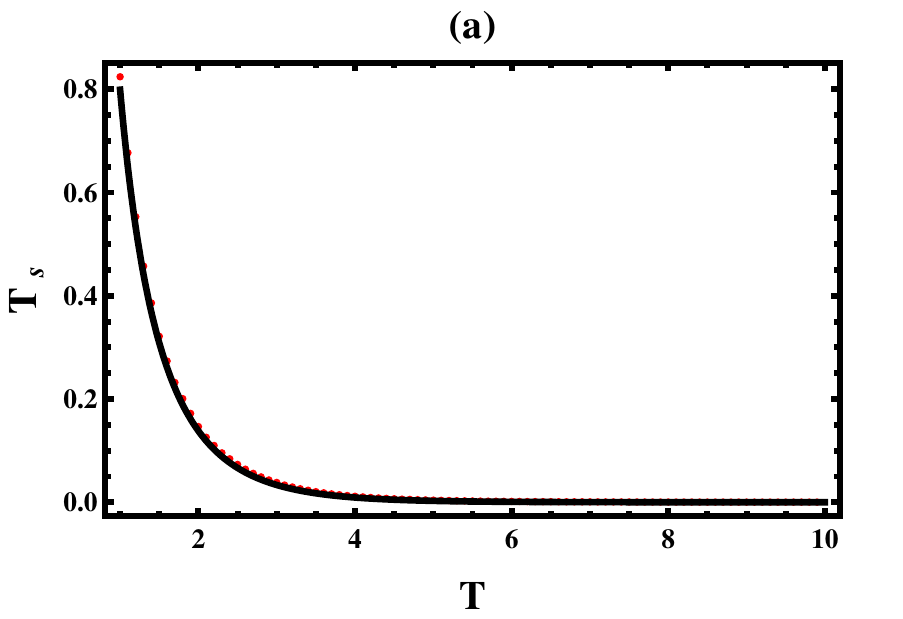}
}
\hspace{1cm}
{
    \includegraphics[width=6cm]{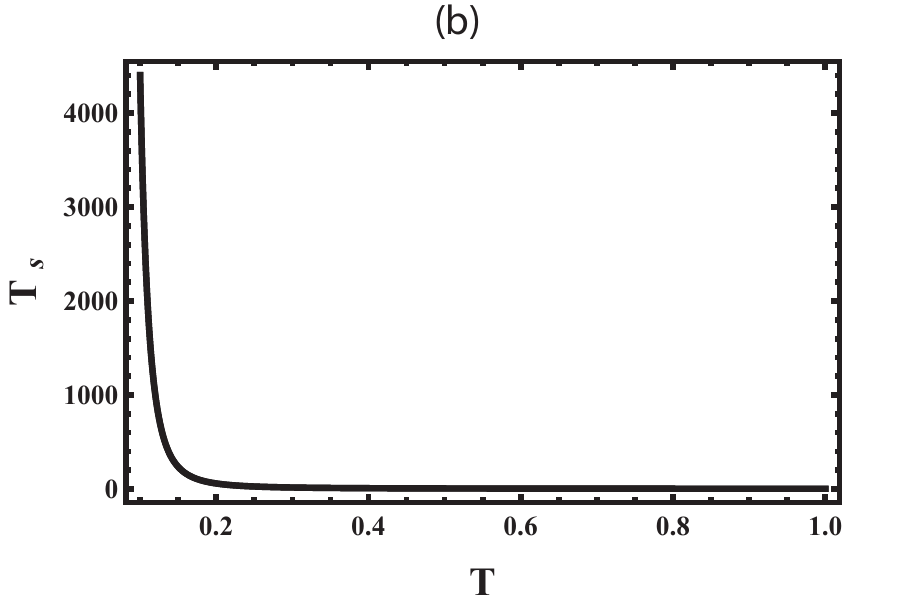}
}
\caption{ (Color online)(a) The mean first passage time $T_s$  as a function of $T$ for the 
parameter values of $ A=1.0$, $U_{0}=1.0$ and $L_{0}=1.0$.   
 (b)  The mean first passage time $T_s$  as a function of $T$ for the  parameter values 
 of $ A=1.0$, $U_{0}=2.0$ and $L_{0}=1.0$. In the figures 
 the red dotted line is evaluated  numerically while the solid line is plotted using 
 the  exact analytic expression (Eq. 10).} 
\label{fig:sub} 
\end{figure}

The exact analytic results  are justified via numerical simulations by integrating the Langevin equation (3) (employing Brownian dynamics simulation).  In the simulation, a Brownian particle
 is initially situated in one of the potential wells. Then the
trajectories for the particle   is simulated by considering different time steps $\Delta t$
and time length $t_{max}$. In order to ensure the numerical accuracy, up to $10^{8}$
ensemble  averages have been obtained

Via numerical simulations  as well as using the exact analytic expression, we 
first plot the MFPT  for temperature dependent viscous friction case ($A=1$) as shown 
in Figs. 2a and 2b. In the figure the red dotted line is evaluated numerically  while
the solid line is plotted using the  exact analytic expression (Eq. 10). The figure depicts that
$T_s$ monotonically decreases as the background temperature increases. In the small regime of $T$ 
(See Fig. 2b), $T_s$ decays exponentially. Exploiting Eq. (10), one can see that the MFPT is 
considerably higher when the viscous friction is temperature dependent ($A=1$) than constant
$\gamma$ case  ($A=0$). As the barrier height increases, $T_{s}$ increases.
\begin{figure}[ht]
\centering
{
    \includegraphics[width=6cm]{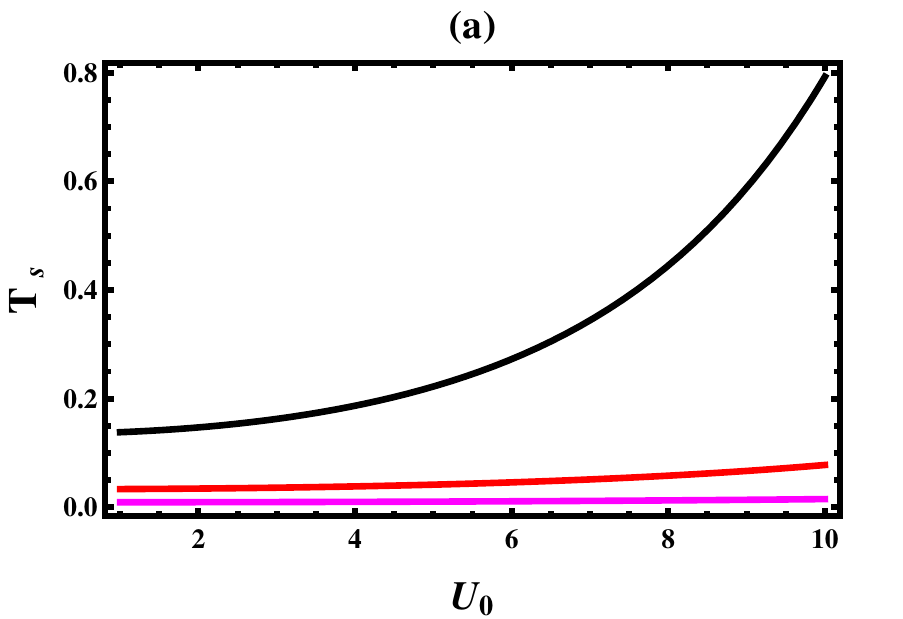}
}
\hspace{1cm}
{
    \includegraphics[width=6cm]{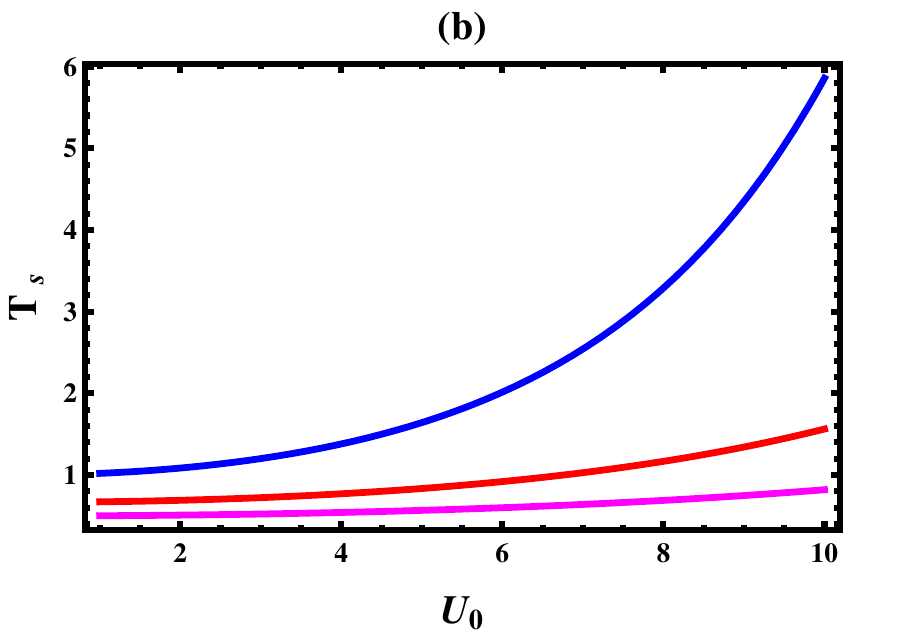}
}
\caption{ (Color online)(a) The mean first passage time $T_s$  as a function of $U_0$ for 
fixed  values of $ A=1.0$ (variable  $\gamma$ case), $T=2.0$, $T=3.0$, $T=4.0$ and $L_{0}=1.0$ from top to bottom, respectively.
 (b)  The mean first passage time $T_s$  as a function of $U_0$ for the  parameter 
 values of $ A=0$ (constant $\gamma$ case), $T=2.0$, $T=3.0$, $T=4.0$ and $L_{0}=1.0$ from top to bottom, respectively.} 
\label{fig:sub} 
\end{figure}

In Figs. 3a and 3b, the mean first passage time $T_s$  is plotted as a function of $U_0$ 
for the  parameter values of $T=2.0$, $T=3.0$, $T=4.0$ and $L_{0}=1.0$ from top to bottom, 
respectively. Fig. 3a represents the constant $\gamma$ while Fig. 3b shows the temperature dependent
$\gamma$ cases.  The figure depicts that $T_s$  decreases monotonically as the background
temperature increases. The same figure depicts also that $T_s$   increases  as the barrier
height $U_0$ steps up.

It is important to note that
most of the previous  studies of thermally activated barrier crossing rate  considered only temperature invariance viscous friction case. In reality, it is well know that the mean first passage time  of a Brownian  particle tends to depend on the intensity of the background temperature. However in liquid or glassy medium,  the viscosity tends to decrease when the intensity of the background temperature increases. This is because  an increase in temperature of the medium brings more agitation to the molecules in the medium, and hence increases their speed. This speedy motion of the molecules creates a reduction in interaction time between neighboring molecules. In turn, at macroscopic level, there will be a reduction in the intermolecular force. Consequently, as the temperature of the viscous medium decreases, the viscous friction in the medium decreases which implies that the mobility  of the particle considerably increases (MFPT decreases) when the temperature of the medium increases. The main message here is that the effect of temperature on the viscous friction is significantly high and cannot be avoided unlike the previous studies.

\subsection{Mean first passage time of many Brownian particles}
Let us now  consider
the First passage time for  one of 
the  $N$ particles  to cross the potential barrier  for the     first  time. Studying such physical problem  is vital and has  been extensively studied in many excitable model systems such as  cardiac systems. Most of these studies have considered  temperature independent viscous friction. In this section we explore further how the temperature of the medium affects the viscosity as well as the the first passage time.

First let us numerically  evaluate the first passage time distribution for a single  and many 
particle systems. This gives us  a qualitative clue on  how the first passage time  behaves 
because the first passage time is given by $T=\int_{0}^{t} t' P_{i}(t')dt'$  where 
$P_{i}(t')$ is the first time distribution of the $i^{th}$ particle.  In Fig. 4, the  first
passage  time distribution of a single particle $P_{i}(t)$  as a function of $t$ is depicted
for  $U_{0}=1.0$ and  $L_{0}=1.0$.  In the figure the $ A=0.0$ (temperature 
independent $\gamma$) and $A=1.0$ (temperature dependent $\gamma$) cases  are shown in the 
red and green lines respectively. Compared to the constant $\gamma$ case, the figure depicts
that the peak of the first time distribution gets higher, and its location shifts to the 
left when the viscous friction is temperature dependent. 
On the other hand, the plot for the first time distribution $P_{N}(t')$ for one of the $N$
particles to fire is  shown in Fig. 5a (constant viscous friction case) and Fig. 5b
(temperature dependent viscous friction case).  As $N$ increases, the peak of the first
passage time distribution decreases revealing that the firing time for one particle
(out of the $N$ particles) decreases as $N$ increases.

\begin{figure}[ht]
\centering
{
    \includegraphics[width=6cm]{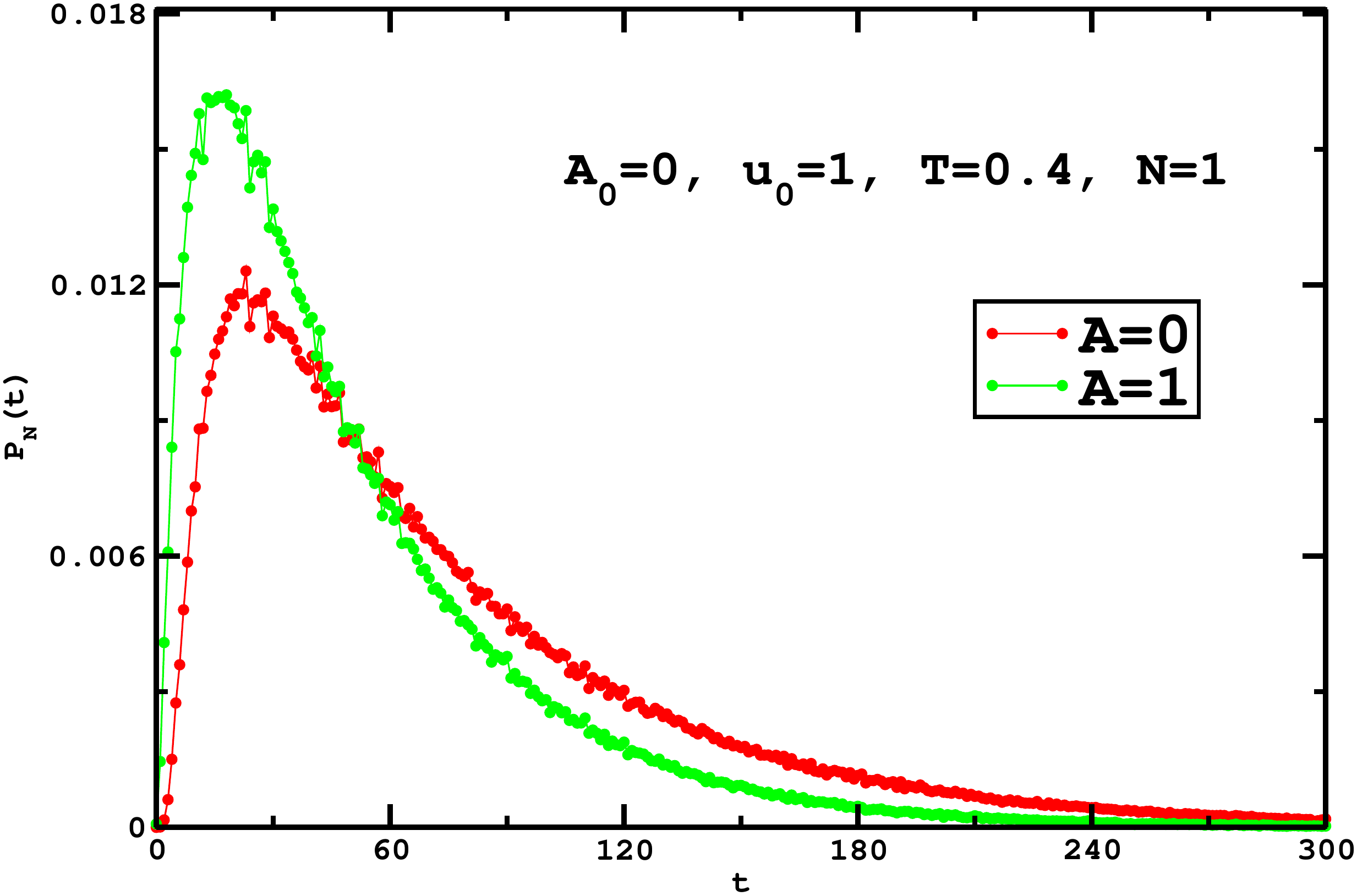}
}
\caption{ (Color online) The first passage time distribution  $P_{i}(t)$  as a 
function of $t$ for $N=1$ and for the  parameter values of $U_{0}=1.0$ and  $L_{0}=1.0$. The plots for  $ A=0$
and $A=1.0$ cases are shown in the red and green lines, respectively.}
\label{fig:sub} 
\end{figure}

\begin{figure}[ht]
\centering
\subfiguretopcaptrue
\subfigure [ ] 
{
    \includegraphics[width=6cm]{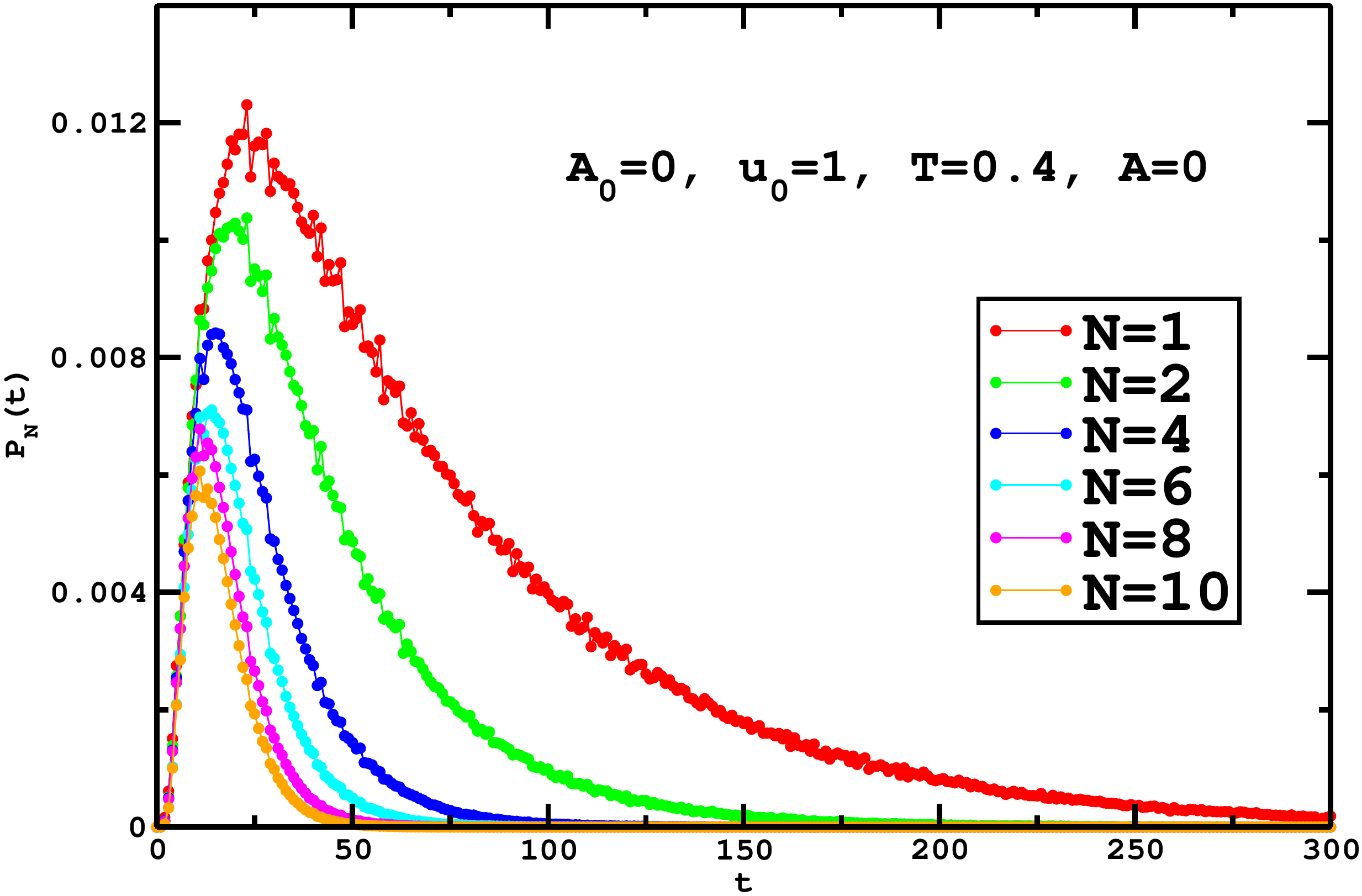}
}
\subfigure [ ] 
{
    \includegraphics[width=6cm]{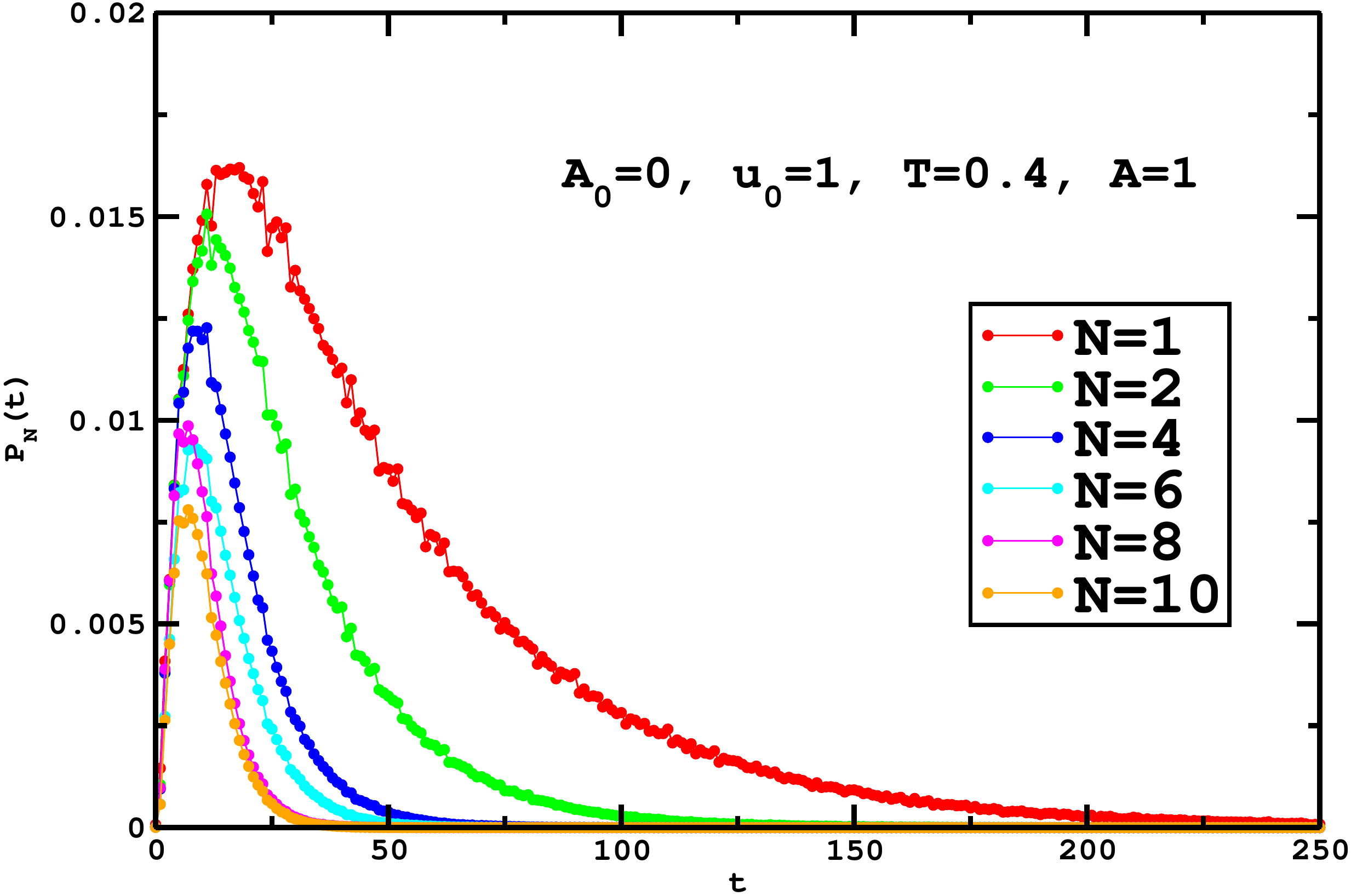}
}
\caption{ (Color online) The first passage time  distribution  $P_{N}(t)$  as a 
function of $t$ for different values of  $N$. We use parameter values of $U_{0}=1.0$ and  $L_{0}=1.0$. Figures (a) 
and (b) show the distributions for the cases $A=0$ and $A=1$, respectively.}
\label{fig:sub} 
\end{figure}

In the high barrier limit, the  first passage time distribution $P_i(t)$ 
is computable as discussed in many litterateurs. To begin with, the Fourier transform of  
first passage time distribution is the characteristic function $\Phi(k,y)$  is given by
\begin{equation} 
\Phi(k,y)=\left\langle exp(ikT)\right\rangle=\sum_{n=0}^{\infty}{(ik)^n\over n!}T_{n}(y).
\end{equation}
Let us define  an integral Kernel $K(y,z')$ as
\begin{equation} 
K(y,z')=\int_y^0 dx {1\over h(x){\mathscr {P}}_s(x)}\int_{-\infty}^{x}dz {\mathscr P}_s(z)\delta(z-z').
\end{equation}
Here $ {\mathscr P}_s(x)$ denotes the equilibrium probability distribution.
Then the characteristic function $\Phi(k,y)$ is derived as 
\begin{eqnarray} 
\Phi(k,y)&=&1+ik \int_{\infty}^{0}dz'K(y,z')+ \nonumber \\
&&(ik)^2\int_{\infty}^{0}dz_{1}\int_{\infty}^{0}dz_{2}K(y,z_1)K(z_1,z_2)+\cdots
\end{eqnarray}
In the high barrier limit, one gets
\begin{equation} 
\Phi(k,y)=\sum_{n=0}^{\infty}{(ikT_s)^n}={i\over i+kT_s}.
\end{equation}
The inverse Fourier transform of $\Phi(k,y)$ is the   first  passage distribution $P_{i}(t)$, and 
after some algebra we get
\begin{equation} 
P_i(t)={e^{{-t\over T_{s}}}\over T_s}
\end{equation}
where $T_s$ is the MFPT for a single particle. 

Once we compute $P_{i}(t)$,  the first passage time distribution  for one particle to cross the barrier out a given $N$ particles  can be evaluated using 
\begin{equation}
P_N(t) = \sum_{i-1}^{N} P_i(t)\prod_{j\neq i}(1-k_j(t))
\end{equation}
where
\begin{equation}
k_j(t)=\int_{0}^{t}dt'P_{j}(t').
\end{equation}
After some algebra we find
\begin{equation} 
P_N(t)={e^{{-t\over T_{N}}}\over T_N}.
\end{equation}
The first arrival time $T_N$, {\em i. e. } the time for one of the particles first to cross the potential barrier,
is calculated via 
\begin{equation} 
T_N=\int_{0}^{t} t' P_{N}(t')dt'.
\end{equation}
For such a case, Eq. (22) reduces to
\begin{equation}
T_N={T_s\over N}={{2TL_0^2e^{-AT}e^{U_0\over T} }\over {NU_0^2}}.
\end{equation}
Exploiting Eq. (23) one can see that as the temperature increases, $T_{N}$ decreases exponentially 
while as the barrier height $U_{0}$ increases, the MFPT decreases. We also note that as the number
of particles increases $T_{N}$ decreases.

 The mean first passage time $T_N$  as a function of $T$ is depicted in Fig. 6a for the 
parameter values of $U_0=2.0$, $N=1.0$, $N=2$,  $N=3.0$ and $L_{0}=1.0$ from top to bottom,
respectively. The viscous friction is considered  to be temperature dependent. In Fig. 6b, 
  the mean first passage time $T_N$  as a function of $T$ is plotted  for the  parameter values of 
 $U_0=2.0$, $N=1.0$, $N=2$,  $N=3.0$ and $L_{0}=1.0$ from top to bottom, respectively  considering  temperature independent viscous friction. As depicted in the figures, $T_{N}$ decreases as the noise strength increases and when the number of particle increases.

\begin{figure}[ht]
\centering
{
    \includegraphics[width=6cm]{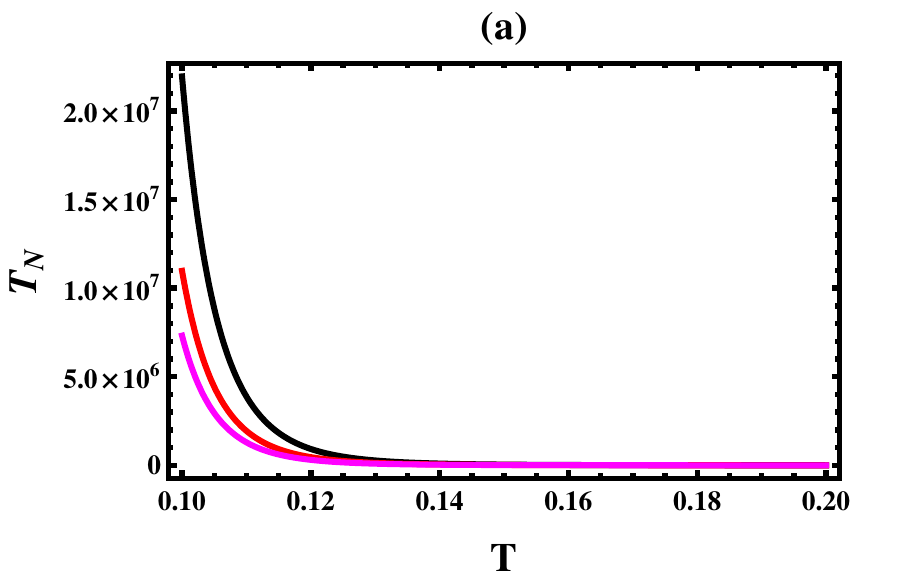}}
{
    \includegraphics[width=6cm]{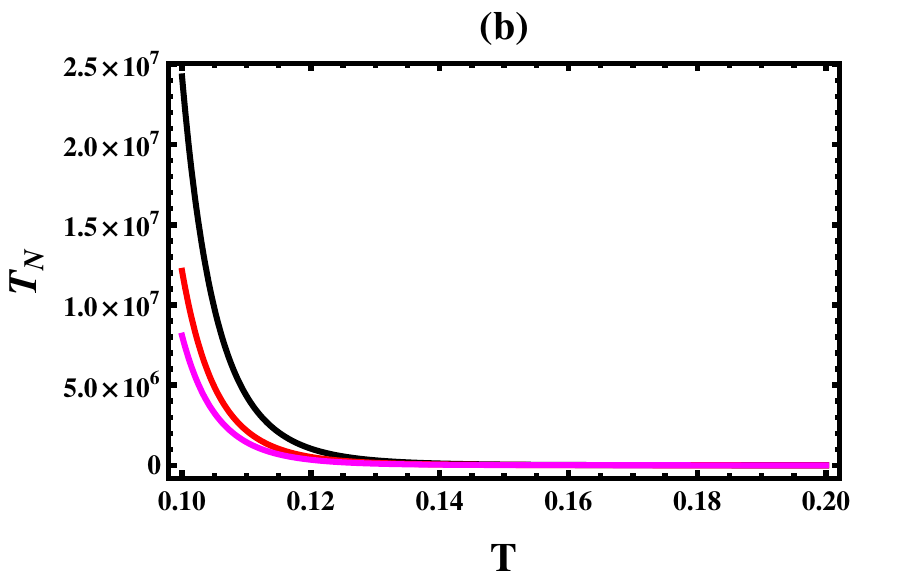}
}
\caption{ (Color online)(a) The mean first passage time $T_N$  as a function of $T$ for the 
parameter values of $U_0=2.0$, $N=1.0$, $N=2$,  $N=3.0$ and $L_{0}=1.0$ from top to bottom,
respectively  for a  variable $\gamma$ case.  
 (b) The mean first passage time $T_N$  as a function of $T$ for the  parameter values of 
 $U_0=2.0$, $N=1.0$, $N=2$,  $N=3.0$ and $L_{0}=1.0$ from top to bottom, respectively  for
 a  constant $\gamma$ case.  } 
\label{fig:sub} 
\end{figure}

\section{Stochastic resonance for a single and many non-interacting particles }  

In the presence of  time varying signal, the interplay between noise and  sinusoidal driving force
in the bistable system may lead the system into stochastic resonance, provided that the random 
tracks are adjusted in an optimal way to the recurring external force. Various studies have used  
different quantities to study the SR of  systems that are driven by a time varying signal. These includes
signal to noise ration (SNR), spectral power amplification ($\eta$), the mean amplitude, as well as 
the residence-time destitution, which all exhibit a pronounced peak at a certain  
noise strength as long as the noise induced hopping events are synchronized with the signal. 
In this section we study the dependence SNR and $\eta$ on the model parameters
by considering a continuous diffusion dynamics and provide a new way to look at the SR on the system.  

In the presence of a time varying periodic signal $A_0 \cos(\Omega t)$, the Langevin equation that governs the 
dynamics of the system is given by 
\begin{eqnarray}
\gamma{dx\over dt}&=&-{\partial U \over \partial x} + A_0 \cos(\Omega t)+ \sqrt{2k_{B}\gamma T}\xi(t).
\label{eqn:sr} 
\end{eqnarray}
where $A_0$ and $\Omega $ are the amplitude and angular frequency of the external signal respectively. 
Eq. 24 is numerically simulated for both  small and large barrier heights. The first
passage time  distribution  $P_i(t)$ shows the resonance profile at the right frequency match.

Before exploring how the signal to noise ratio as well as spectral amplification behaves on the 
model parameters, first let us explore the dependence of the first passage time distribution 
on system parameters  numerically  by integrating Eq. 24. Figure  (7) shows the first time 
distribution function $P_N(t)$  as a function of time for $U_{0}=1.0$, 
$T=0.4$ and $A=0$. In Figs. 7a, 7b and 7c, the number of particles is fixed at $N=1$, $N=4$ and 
$N=8$, respectively. To compare with, we have plotted the distributions  both in the presence
of signal $A_0=1.0$ (red solid line) and in the absence of signal $A_0=0.0$ (green solid line). 
Only in the presence of signal that the the distributions shows the points of resonances.  As the number of particles increase the number 
of local maxima fades out. 
\begin{figure}[ht]
\centering
\subfiguretopcaptrue
\subfigure [ ] 
{
    \includegraphics[width=6cm]{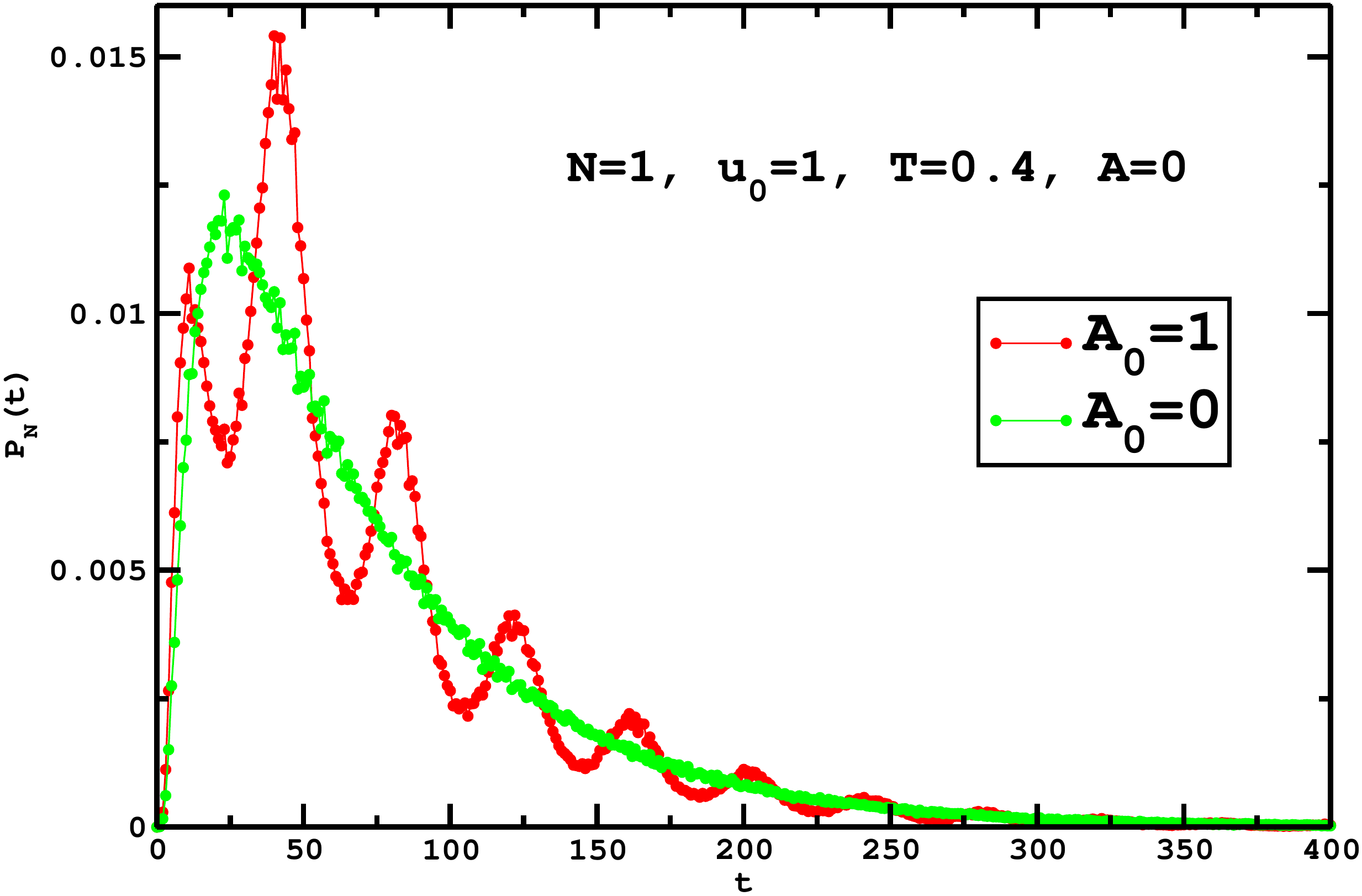}
}
\subfigure[ ] 
{
    \includegraphics[width=6cm]{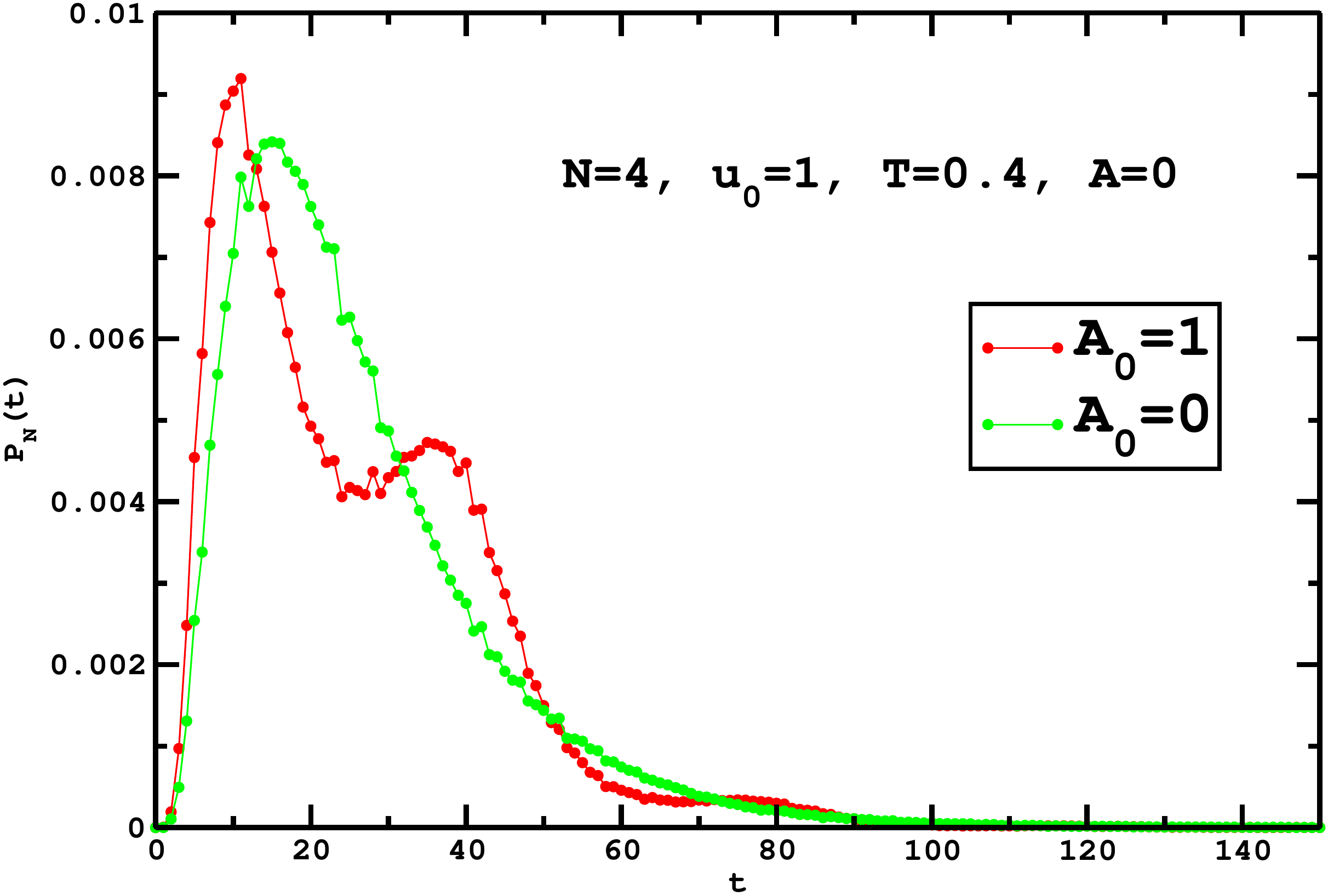}
}
\subfigure [ ] 
{
    \includegraphics[width=6cm]{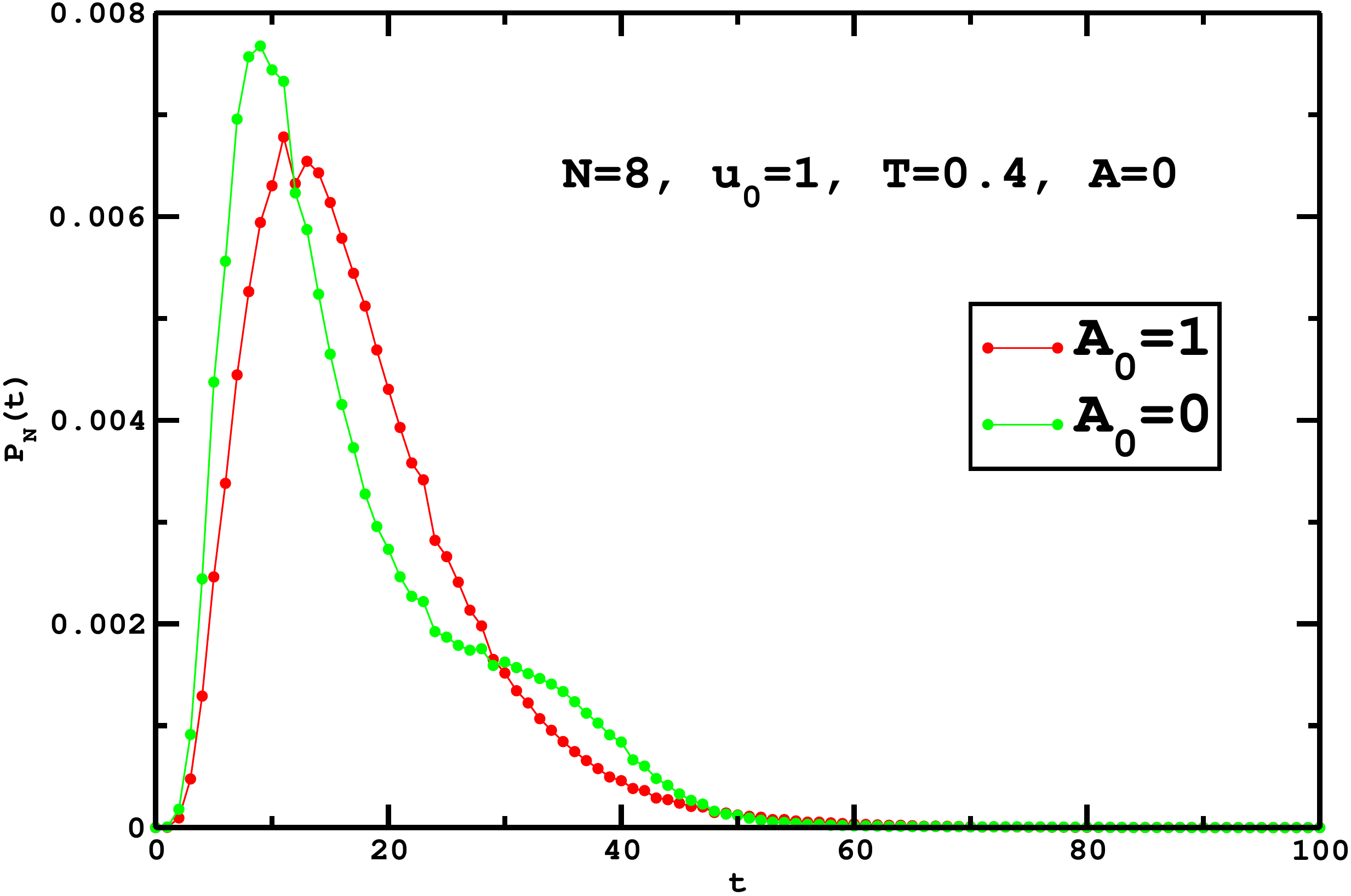}
}

\caption{ (Color online) The first passage time  distribution  $P_N(t)$  as a function of time 
for the parameter values of $U_0=1.0$, and $A=0$.  The red and green lines are plotted when the
external signal is turned on and off, respectively. In Figs. (a),(b) and (c),  $N$ is fixed 
as $N=1$, $N=4$ and $N=8$, respectively.} 
\label{fig:sub} 
\end{figure}
The resonance profile can be observed better by looking at the relative ratios of the first passage time 
distribution functions with and without external periodic signal, {\em i. e.} taking the ratios of green
and red lines in Fig. 8. It turned out that the ratio of the distribution is independent of the number 
of particles in the system as shown in Fig. 8a for single particle case and Fig. 8b for many particles case.
\begin{figure}[ht]
\centering
\subfiguretopcaptrue
\subfigure[ ] 
{
    \label{fig:sub:a}
    \includegraphics[width=6cm]{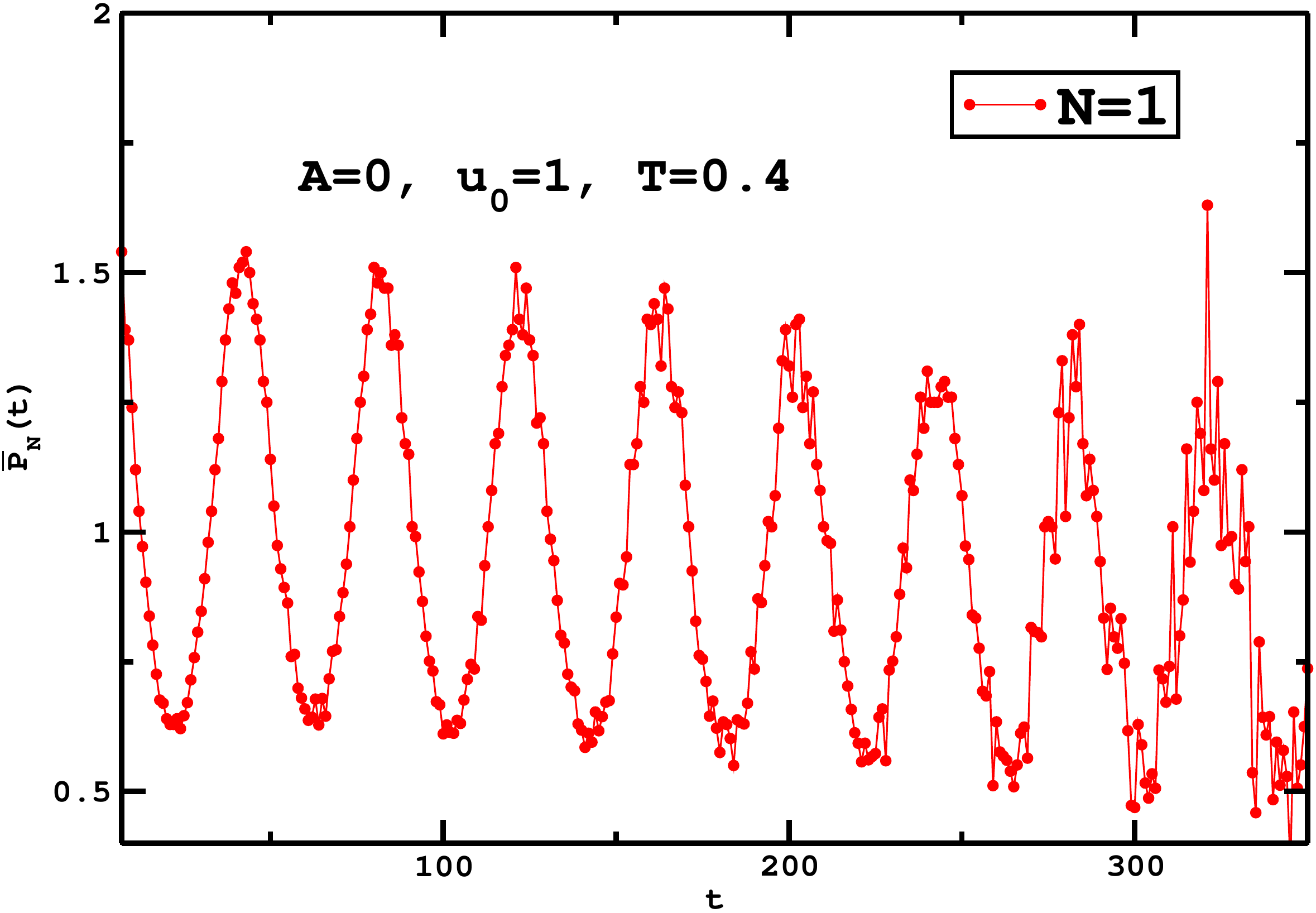}
}
\subfigure[ ] 
{
    \label{fig:sub:b}
    \includegraphics[width=6cm]{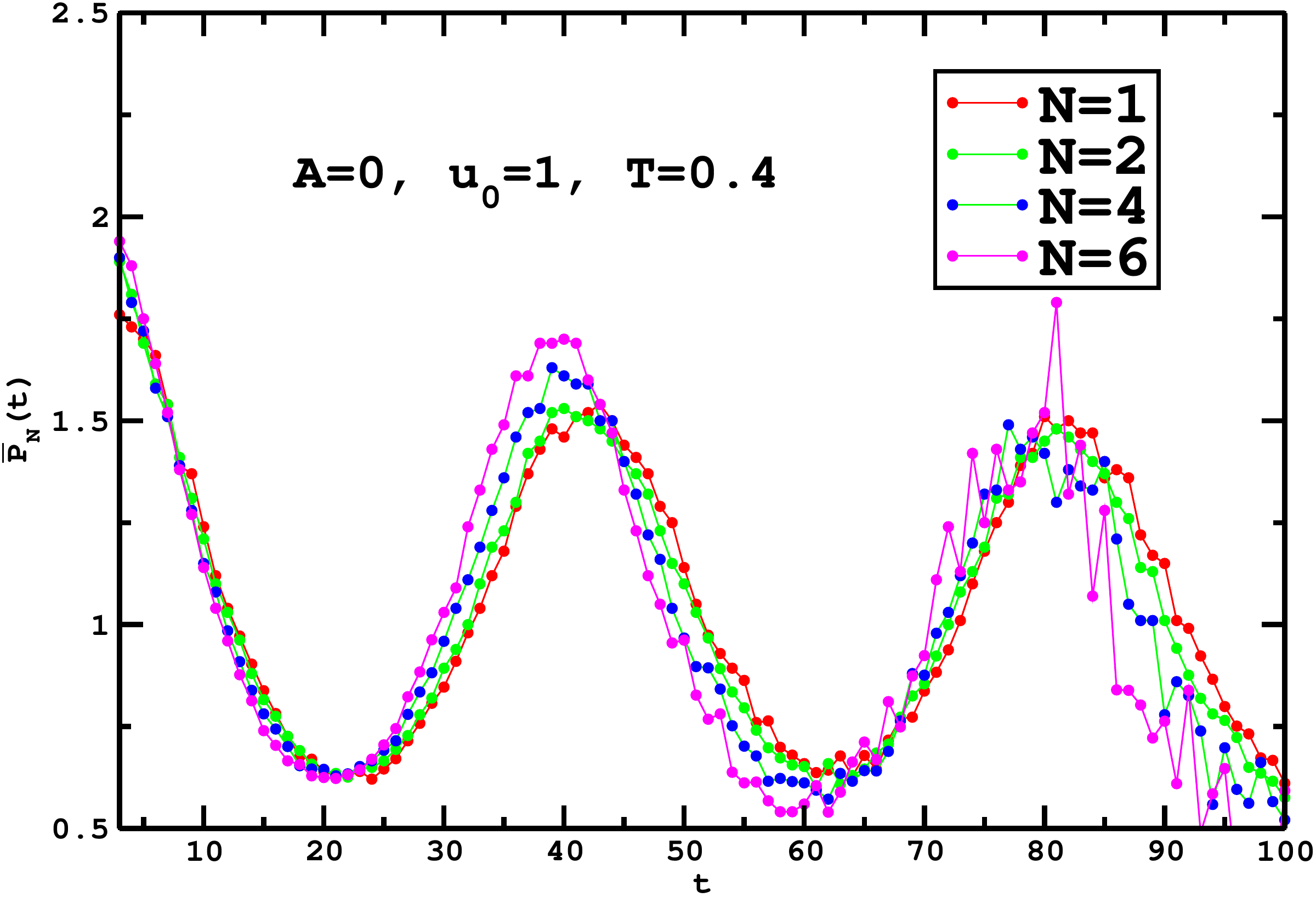}
}

\caption{ (Color online) The ratio of the first passage time distribution functions $\bar{P}_{N}(t)$. 
(a) The number of particle is fixed as  $N=1$. (b) In the figure, $N$ is fixed as $N=1$, $N=2$,
$N=4$ and $N=6$.} 
\label{fig:sub} 
\end{figure}

\begin{figure}[ht]
\centering
\subfiguretopcaptrue
\subfigure[ ] 
{
    \includegraphics[width=6cm]{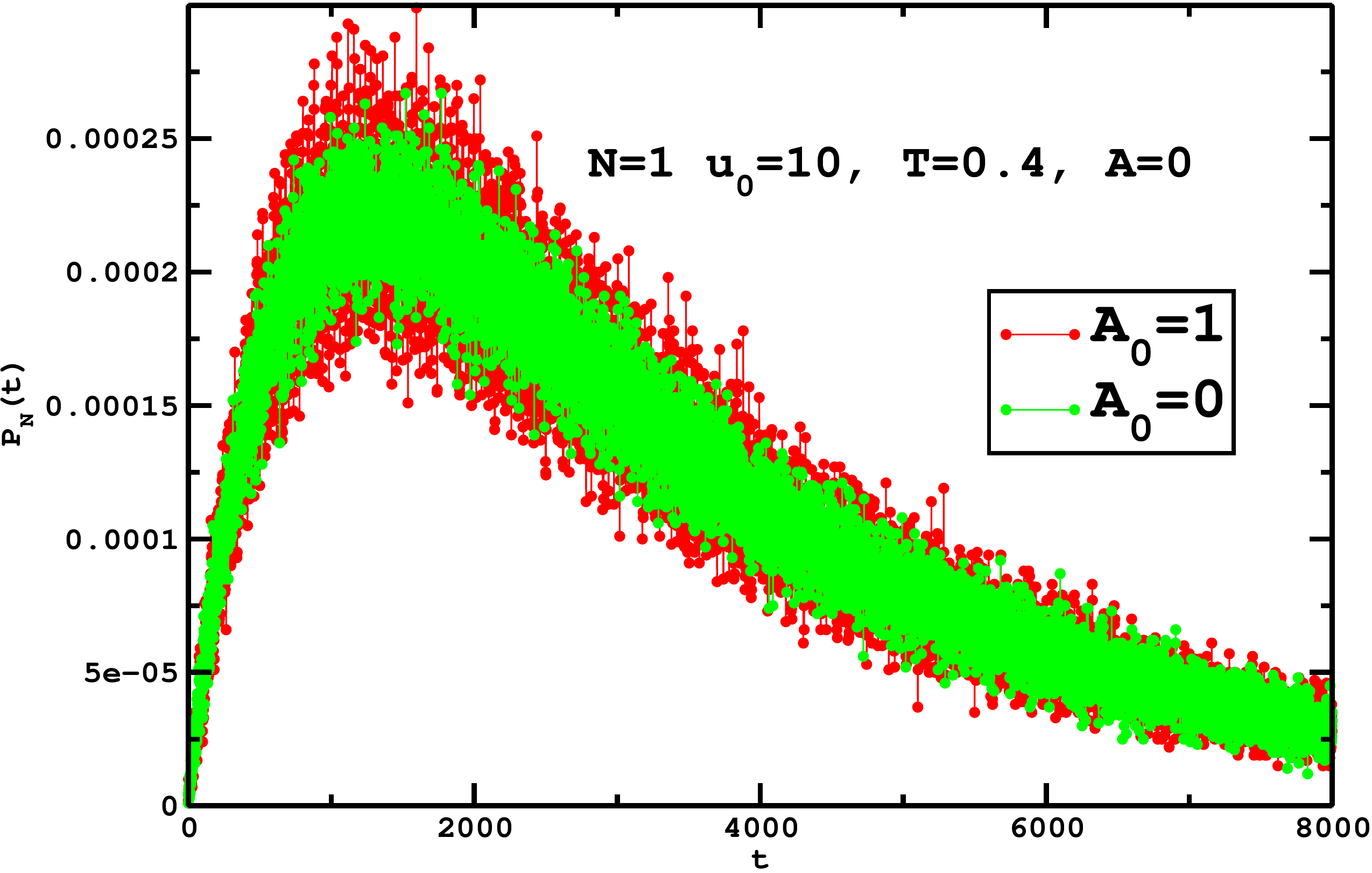}
}
\subfigure[ ] 
{
    \includegraphics[width=6cm]{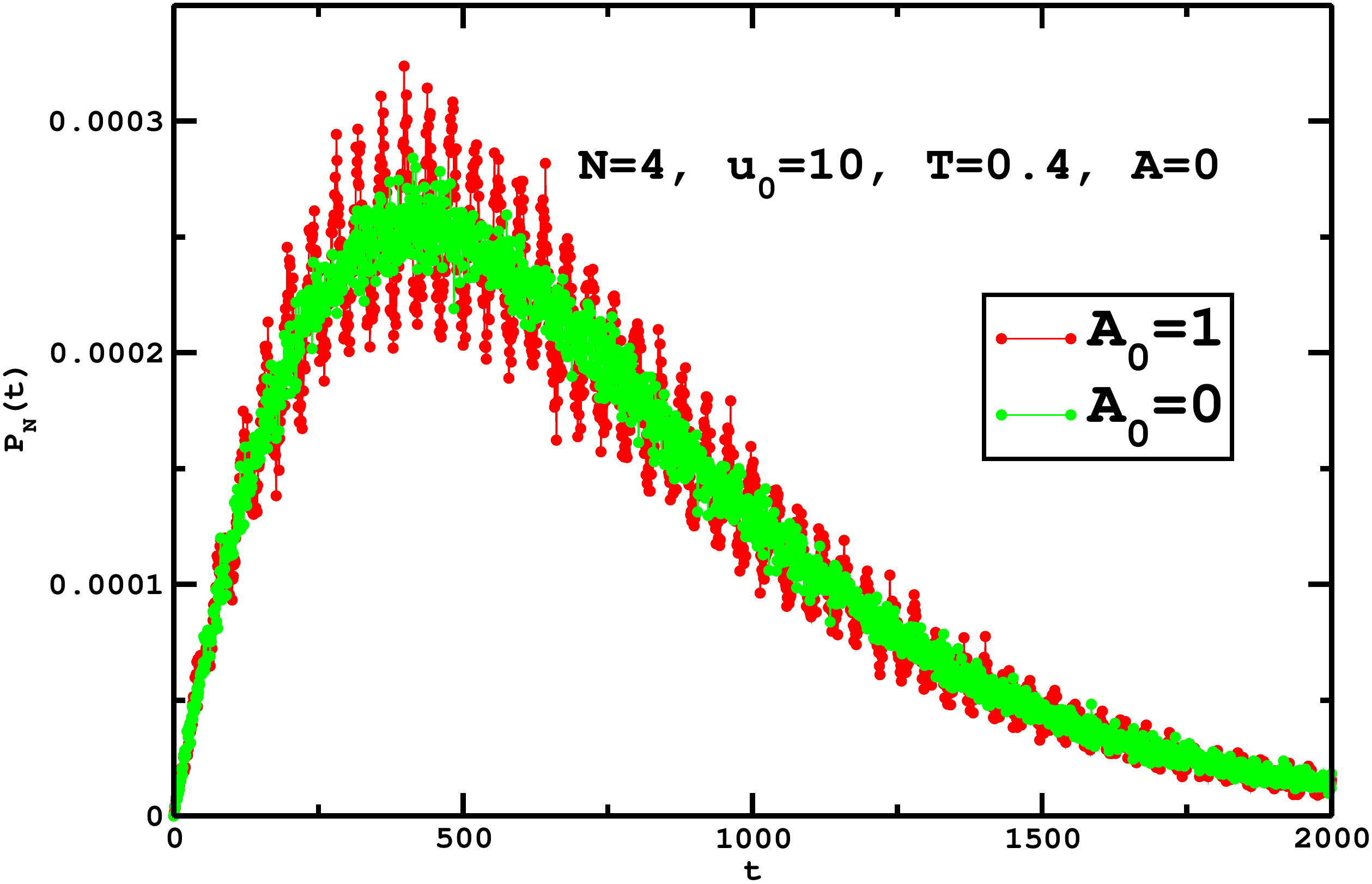}
}
\subfigure[ ] 
{
    \includegraphics[width=6cm]{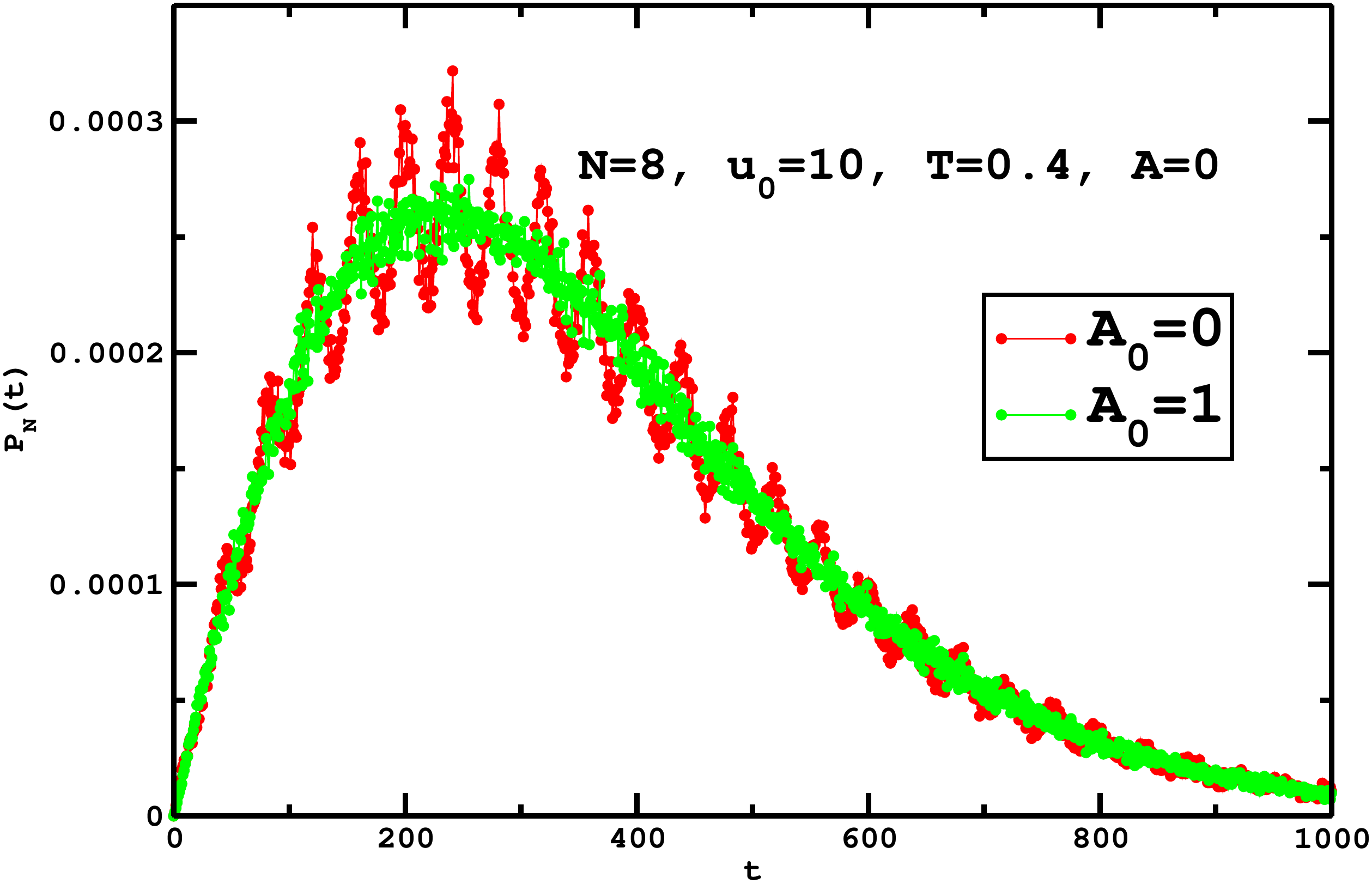}
}
\caption{ (Color online) The  first passage time 
distribution $P_N(t)$  as a function of time 
in high  barrier limit. In the figure, the parameters are fixed as  $U_0=10.0$ and $A=0$. Figs. (a), (b) and (c) are plotted  by considering 
 one, four and eight particles, respectively.
} 
\label{fig:sub} 
\end{figure}

In high barrier limit we see more peaks. In Figs. 9a, 9b and 9c, we plot the first passage time 
distribution time in high barrier limit. In the figures,  the number of particles is fixed as
$N=1$, $N=4$ and $N=8$, respectively. To compare with, we have plotted the distributions  both
in the presence of signal $A_0=1.0$ (red solid line) and in the absence of signal $A_0=0.0$ 
(green solid line). Only in the presence of signal that the  the distributions shows the points 
of resonances.

To observe more the effect of the temperature dependence  of $\gamma$ on the SR we have plotted the first passage time
distribution functions in the presence of external force when $A=0$ and $A= 1$ in the limit of small 
barrier height  as shown in Fig. 10. In Figs. 10a, 10b and 10c, the number of particles is fixed as $N=1$, $N=4$ and $N=8$, respectively. 
The figures shows that the resonance is more pronounced when $\gamma$ is
temperature dependent.

\begin{figure}[ht]
\centering
\subfiguretopcaptrue
\subfigure[] 
{
    \includegraphics[width=6cm]{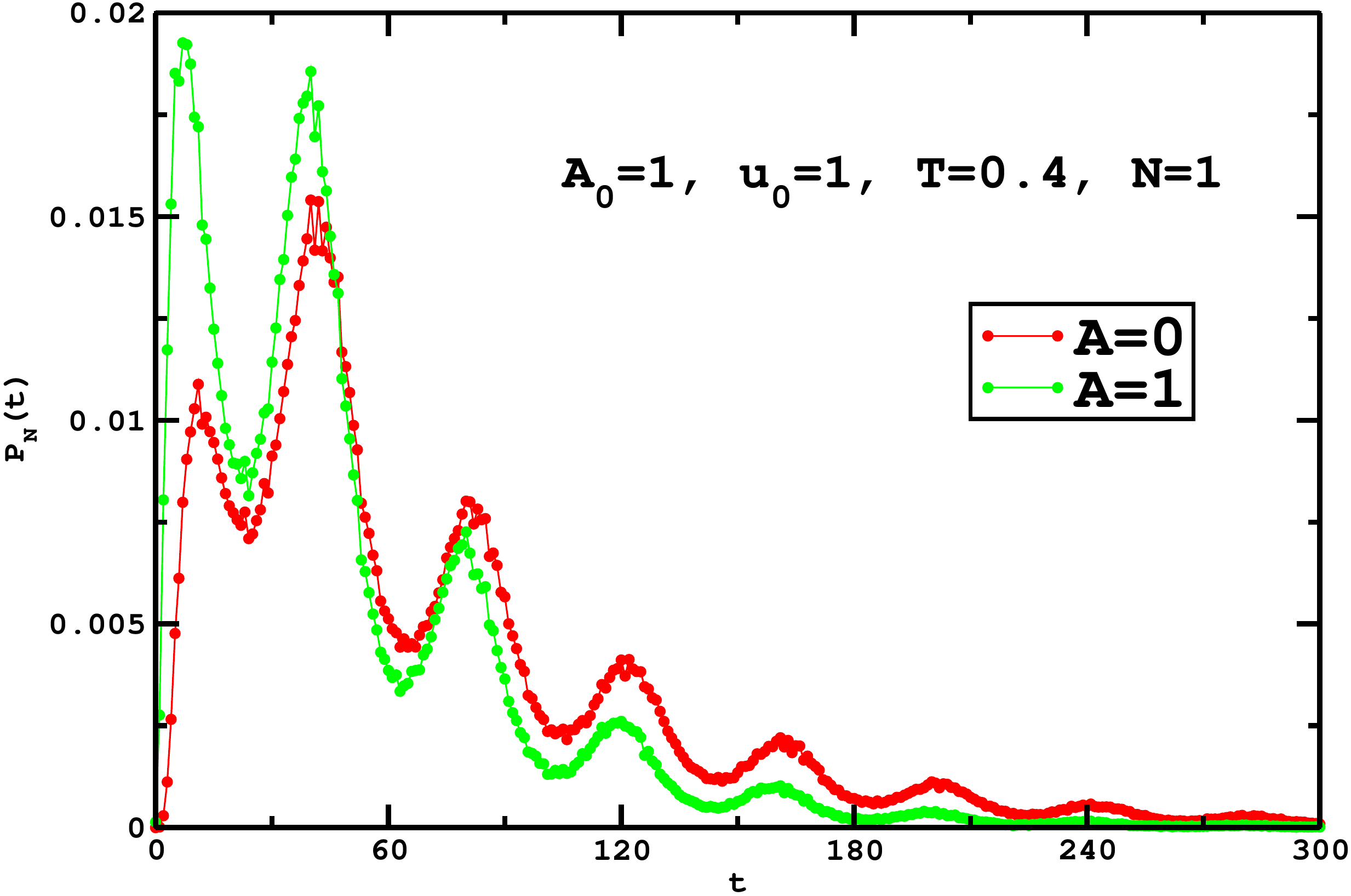}
}
\subfigure[] 
{
    \includegraphics[width=6cm]{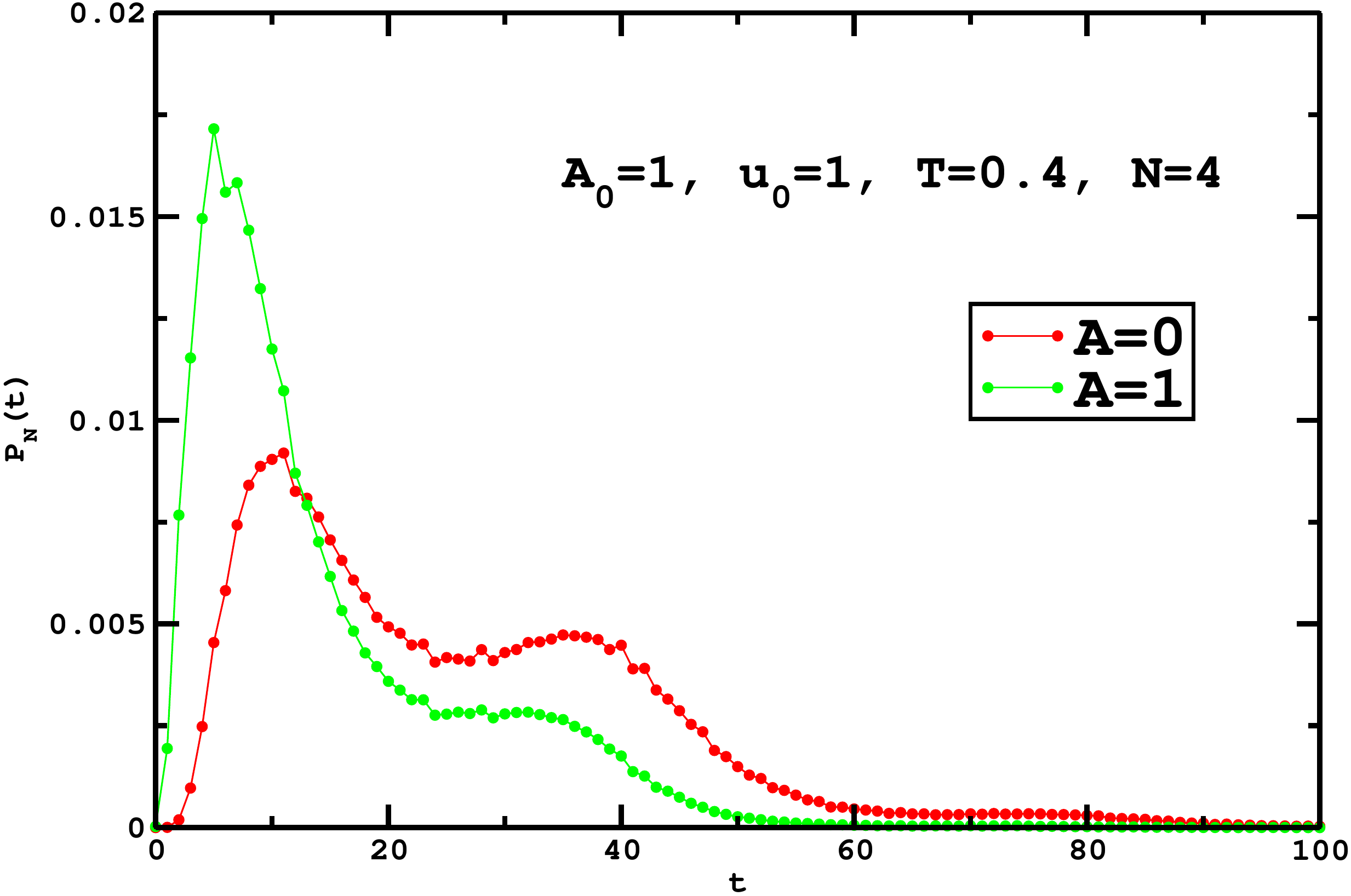}
}
\subfigure[] 
{
    \includegraphics[width=6cm]{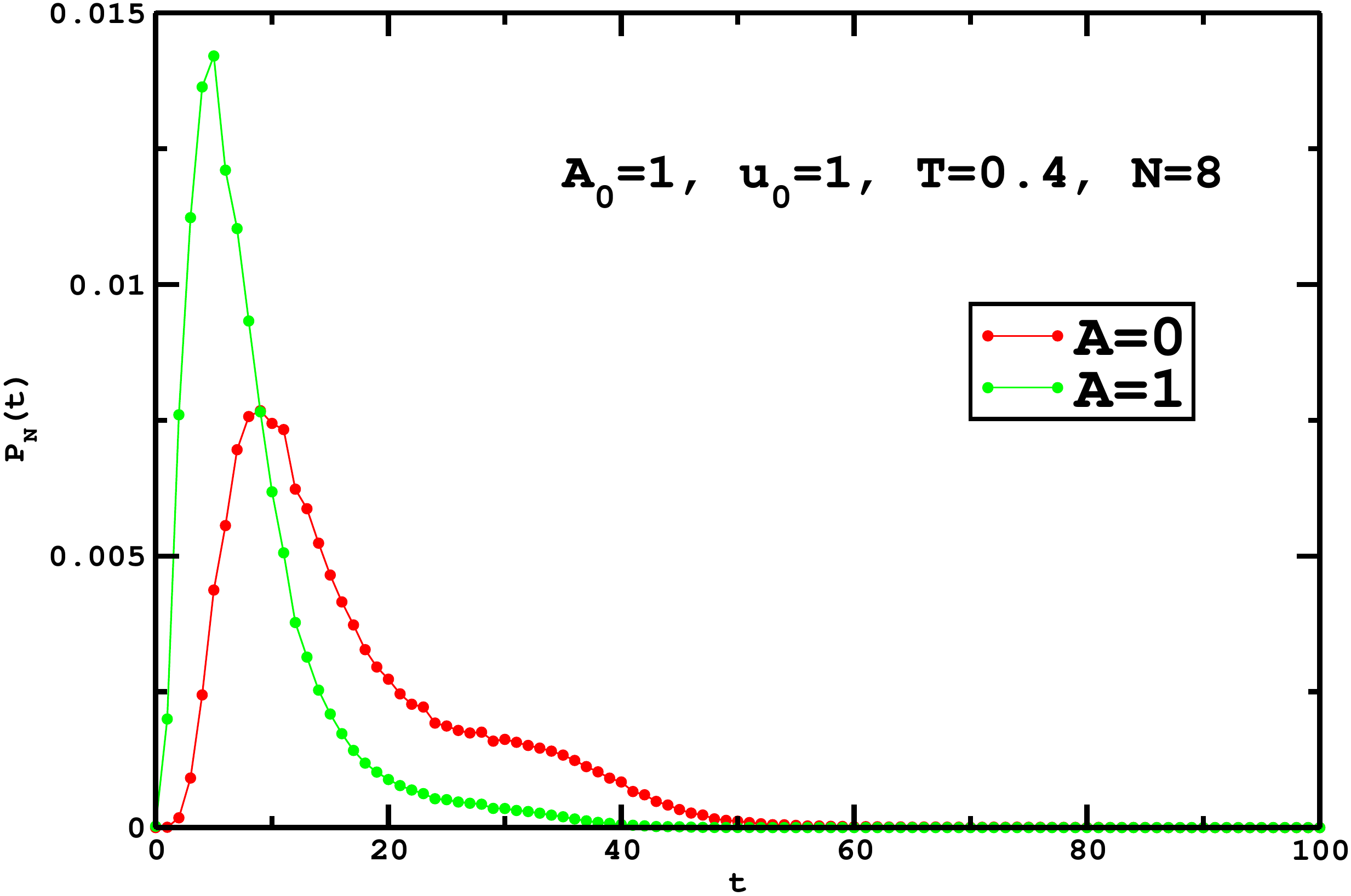}
}

\caption{ (Color online) The  first passage time 
distribution $P_i(t)$  as a function of time 
in small  barrier limit. In the figure, the parameters are fixed as  $U_0=10.0$ and $A=0$. Figs. (a), (b) and (c) are plotted  by considering 
 one, four and eight particles, respectively.
} 
\label{fig:sub} 
\end{figure}

\subsection{ Signal to noise ratio}

The signal to noise ratio can be studied via two state model. 
Employing two state model approach \cite{am14}, two discrete states $x(t) =\pm L_0$ are considered.
Let us denote $n_+$ and $n_{-}$ to be the probability to find the particle in the right 
($L_0$) and in the left ($−L_0$) sides of the potential wells, respectively. In the presence 
time varying signal, the master equation that governs the time evolution of $n_{\pm}$ is given by
\begin{eqnarray}
 {\dot n}_{\pm}(t) = −W_{\pm}(t)n_{\pm} + W_{\mp}(t)n_{\mp}
\end{eqnarray}
where $W_{+}(t)$ and $W_{-}(t)$ corresponds to the time dependent transition probability 
towards the right ($L_0$) and the left (−$L_0$) sides of the potential wells. The time 
dependent rate \cite{am14} takes a simple form 
\begin{eqnarray}
W_{\pm}(t) = R \exp\left[\pm{ L_0  A_0\over U_0 T} \cos(\Omega t)\right]
\end{eqnarray}
 where $R$ is the Kramers rate for the particle in the absence of periodic force $A_0 = 0$. 
 For sufficiently small amplitude, one finds the signal to noise ratio to be
\begin{eqnarray}
SNR =N \pi R\left({ L_0  A_0 e^{-T}\over U_0 T}\right)^2
\end{eqnarray}
when $\gamma$ is temperature dependent and
\begin{eqnarray}
SNR = N\pi R\left({ L_0  A_0\over U_0 T}\right)^2
\end{eqnarray}
when $\gamma$ is constant. Here the rate can be found by substituting 
\begin{eqnarray}
R=N/T_{s}.
\end{eqnarray}
\begin{figure}[ht]
\centering
{
    \includegraphics[width=6cm]{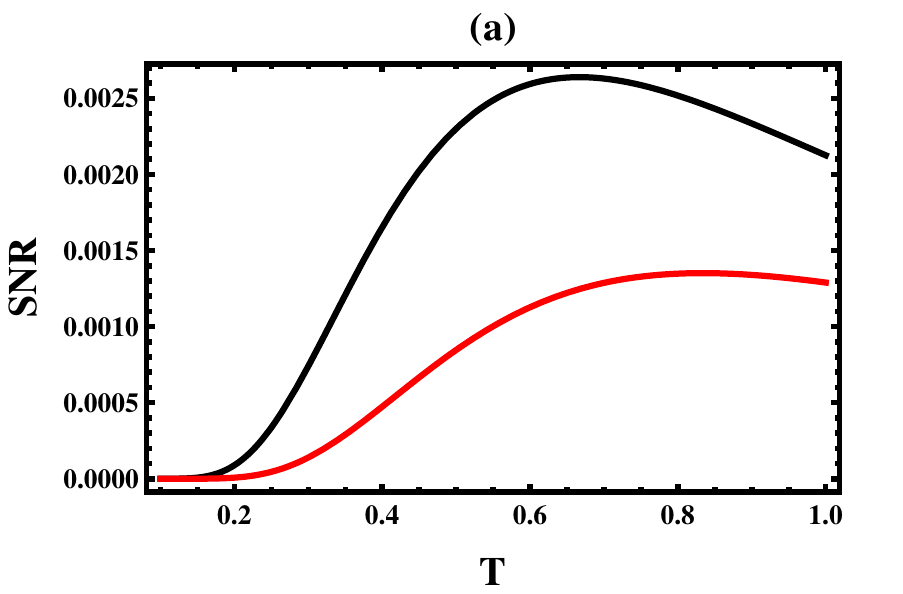}}
\hspace{1cm}
{
    \includegraphics[width=6cm]{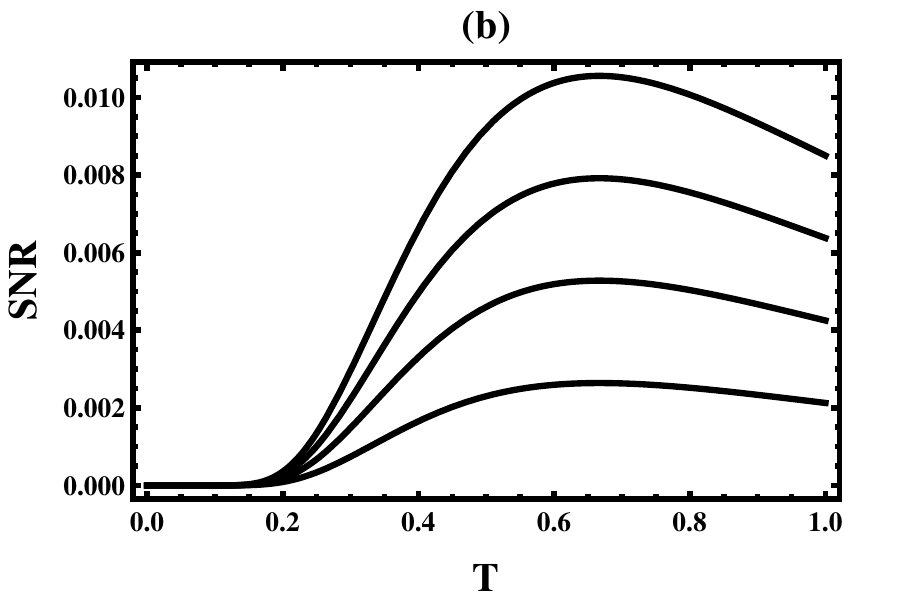}
}
\caption{ (Color online)(a) The  $SNR$  as a function of $T$ for the  parameter values of
$A_0=0.1$, $U_0=2.5$, $U_0=2.0$ and $L_{0}=1.0$ from top to bottom, respectively  for a  variable 
gamma case.  
 (b) The $SNR$   as a function of $T$ for the  parameter values of $A_0=0.1$, $U_0=2.0$, $N=1.0$, 
 $N=2$,  $N=3.0$, $N=4$ and $L_{0}=1.0$ from top to bottom, respectively  for a  
 constant gamma case.  } 
\label{fig:sub} 
\end{figure}

Before we explore how the SNR  behaves as a function of $N$, we introduce
additional dimensionless parameter: ${\bar A_0}= A_0 L_0/U_0$, and for brevity
we drop the bar hereafter. Fig. 11a depicts the plot for the SNR as a function of $T$ for
the  parameter values of $A_0=0.1$, $U_0=2.5$ and $U_0=2.0$ and $L_{0}=1.0$ from top to bottom, 
respectively  for a  variable gamma case.   The SNR exhibits monotonous noise strength dependence revealing a peak 
at an optimal noise strength $T_{opt}$. $T_{opt}$ steps down as $A_0$  decreases.
In Fig. 11b, the SNR as a function of $T$ is plotted  for the  parameter values of
$U_0=2.0$, $N=1.0$, $N=2$,  $N=3.0$, $N=4$ and $L_{0}=1.0$ from   bottom to top, 
respectively  for a  constant gamma case and 
$A_0=0.1$. As shown in the figures the SNR increases with $N$.

\subsection { The power amplification factor} 

To gain  more  understanding of the SR of the Brownian particle, we  consider the linear
response of the particle   to the small driving forces. Following the same approach as
our previous work  \cite{am24}, in the linear response  regime, we find  the power 
amplification power as 
\begin{eqnarray}
\eta = \left( {\langle X^2 \rangle} \over T \right)^2 { {4R^2} \over {4R^2+\Omega^2}}
\end{eqnarray}
where $\left\langle X^2\right\rangle=\int X^2e^{{-U_0\over T}}dX/ \int e^{{-U_0\over T}}dX$.
In our case after some algebra we find 
\begin{eqnarray}
\left\langle X^2\right\rangle={L_0^2 (-2 T^2 + e^{U_0/T} (2 T^2 - 2 T U_0 + U_0^2))
\over  {U_0^2(-1 + e^{U_0/T})}}
\end{eqnarray}
and as usual the rate $R=N/T_{s}$ where $T_{s}$ is given by  Eq. (10) (variable $\gamma$) or
Eq. (12) (constant $\gamma$).
\begin{figure}[ht]
\centering
{
    \includegraphics[width=6cm]{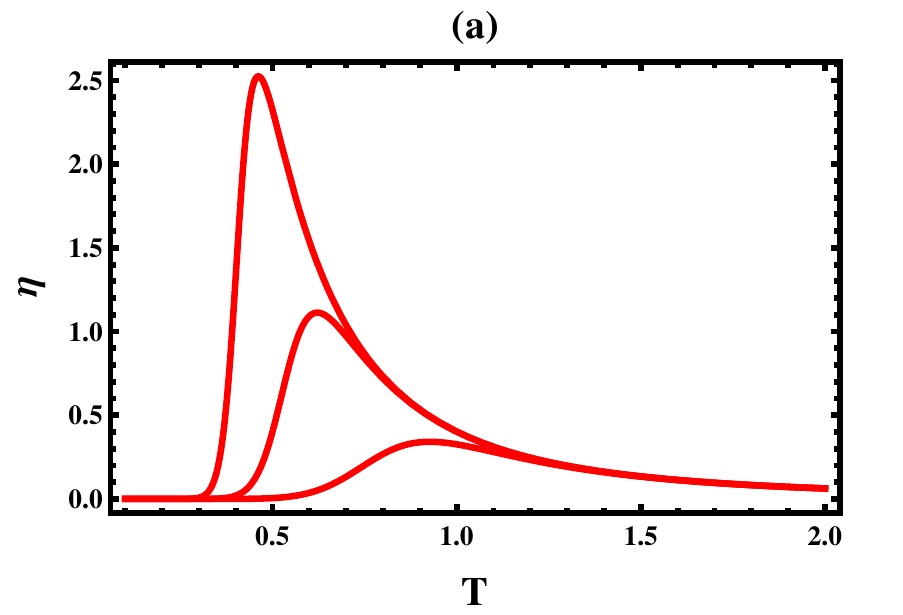}}
\hspace{1cm}
{
    \includegraphics[width=6cm]{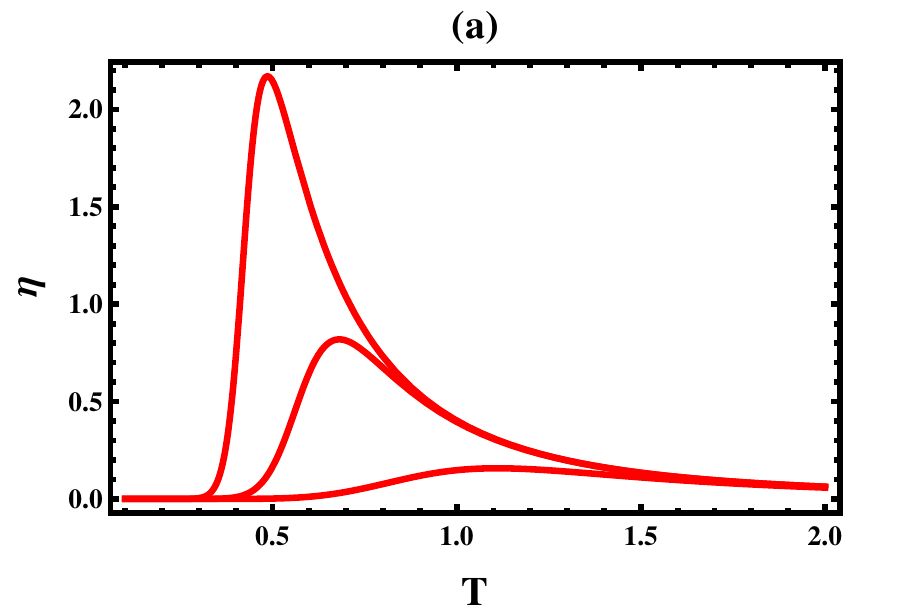}
}
\caption{ (Color online)(a)   $\eta$  as a function of $T$ for the  parameter values of $U_0=4$,
$\Omega=0.004$, $\Omega=0.04$ and $\Omega=0.4$  from top to bottom, respectively  for a 
variable $\gamma$ case.  
 (b)  $\eta$  as a function of $T$ for the  parameter values of $U_0=4$, $\Omega=0.004$,
 $\Omega=0.04$ and $\Omega=0.4$  from top to bottom, respectively  for   a  constant 
 $\gamma$ case.  } 
\label{fig:sub} 
\end{figure}

The spectral amplification  $\eta$  as a function of $T$  is plotted in Fig. 12a for the 
parameter values of $U_0=4$, $\Omega=0.004$, $\Omega=0.04$ and $\Omega=0.4$  from top to 
bottom, respectively  for a  variable $\gamma$ case.  The figure depicts that $\eta$ exhibits a pronounced peak at a particular 
$T_{opt}$. As $\Omega$ increases $\eta$ steps down and $T_{opt}$ shifts to the right. 
This is reasonable since resonance occurs when  ${1\over T_s}=\pi\Omega$. As $\Omega$ 
steps up, $T_{s}$ should decreases in order to keep the resonance condition. However $T_{s}$  
decreases only when  $T$ increases.
 In Fig. 12b,  we plot $\eta$  as a function of $T$ for the  parameter values of $U_0=4$, 
 $\Omega=0.004$, $\Omega=0.04$ and $\Omega=0.4$  from top to bottom, respectively  for 
 a  constant $\gamma$ case.  
The same figures exhibits that the SNR is considerably lower  for temperature  dependent viscous friction case.

On the other hand for many particle cases   $\eta$  as a function of $T$ is depicted 
in Fig. 13a for the  parameter values of $U_0=4.0$, $\Omega=0.004$, $N=10$, $N=5$ and
$N=1$  from top to bottom, respectively  for a  variable $\gamma$ case.   The figure clearly  exhibits that $\eta$ steps
up as $N$ increases. The figure depicts that $\eta$ exhibits a pronounced peak at 
a particular $T_{opt}$. As $N$ increases $\eta$ steps up and $T_{opt}$ shifts to the
left. This is plausible since resonance occurs when  ${N\over T_s}=\pi\Omega$. Here 
since $\Omega$ is fixed, as $N$ increases, $T_s$ should increases to obey the resonance
condition.  However $T_{s}$  increases,  only when  $T$ decreases.
 In Fig 13b we plot  $\eta$  as a function of $T$ for the  parameter values of $U_0=4.0$,
 $\Omega=0.004$, $N=10$, $N=5$ and $N=1$  from top to bottom, respectively  for a 
 constant $\gamma$ case.   The
 same figure exhibits that the peak of $\eta$ is smaller  in comparison to that of  variable $\gamma$ case.
\begin{figure}[ht]
\centering
{
    \includegraphics[width=6cm]{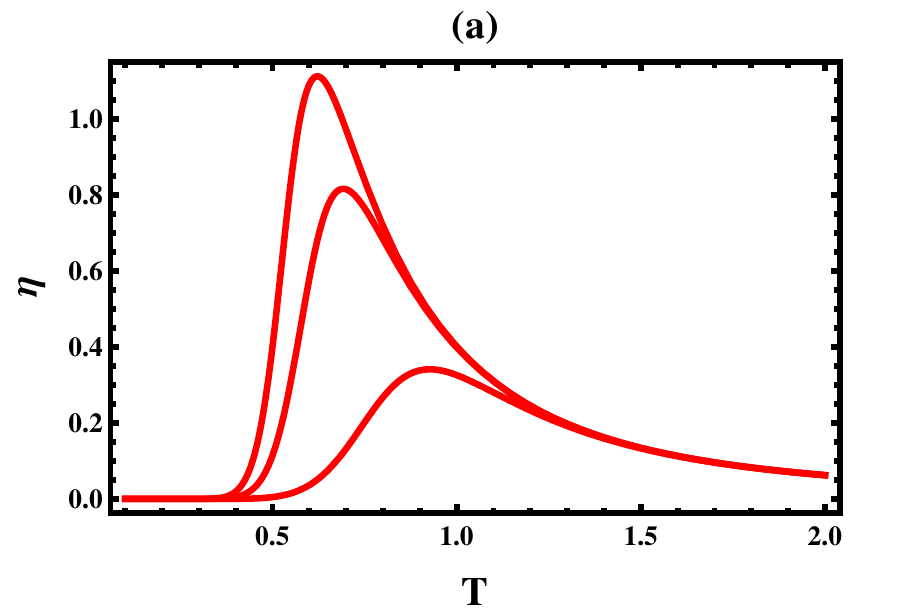}}
\hspace{1cm}
{
    \includegraphics[width=6cm]{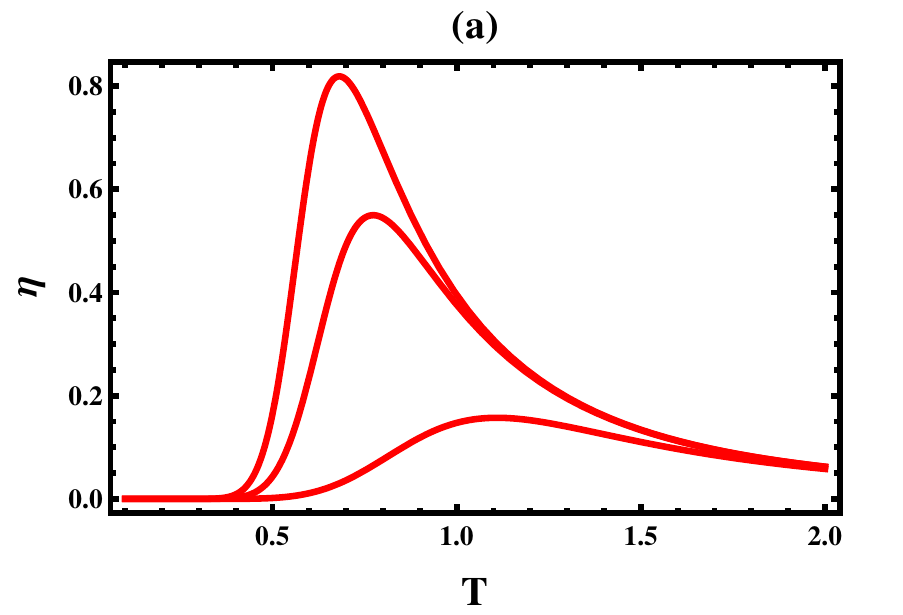}
}
\caption{ (Color online)(a)  $\eta$  as a function of $T$ for the  parameter values of $U_0=4.0$, 
$\Omega=0.004$, $N=10$, $N=5$ and $N=1$  from top to bottom, respectively  for a  variable $\gamma$ 
case.  
 (b) $\eta$  as a function of $T$ for the  parameter values of $U_0=4.0$, $\Omega=0.004$, 
 $N=10$, $N=5$ and $N=1$  from top to bottom, respectively  for a  constant $\gamma$ case.  
} 
\label{fig:sub} 
\end{figure}

\section {Summary and conclusion}
 
In the present  work, a generic model system is presented which helps to understand the dynamics 
of excitable systems such neural and cardiovascular systems.
 The role of noise on the first passage time is investigated in detailed. Particularly, the role 
 of temperature on the viscous  friction as well as on the MFPT  is explored by considering  
a viscous friction $\gamma$ that  decreases exponentially when the temperature $T$ of the
medium increases  ($\gamma=Be^{-A }$) as proposed originally by Reynolds \cite{am10}. We 
show  that the MFPT is smaller in magnitude when $\gamma$ is temperature dependent than 
temperature independent $\gamma$ case which is reasonable  because  the diffusion constant
$D=T/\gamma= k_{B}T e^T$  is valid when   viscous friction to be temperature dependent showing that 
the effect of temperature on the particle mobility is considerably high.

In this work first we study the MFPT of a single particle both for temperature dependent 
and independent viscous friction cases. The exact analytic result as well as the simulation 
results depict that the MFPT is considerably smaller when $\gamma$ is temperature dependent. 
In both cases the escape rate increases as the noise strength increases and  decreases as
the potential barrier increases.  We then  extend our study for $N$ particle systems. The 
first passage time $T_{N}$ for one particle out of $N$ particles to cross the potential
barrier can be studied both analytically  at least in the high barrier limit and via 
numerical simulation for any cases. It is found that 
$T_{N}$ is considerably  smaller when the viscous friction is temperature dependent. 
For both cases, $T_{N}$ decreases as the noise strength increases and as the potential 
barrier steps down.  In high barrier limit, $T_{N}=T_{s}/N$  where $T_{s}$ is the MFPT 
for a single particle.  In general as the number of particles  increases, $T_{N}$ decreases.

We then study our  model system in the presence of time varying signal. In this case 
the interplay between noise and sinusoidal driving force in the bistable system may 
lead the system into stochastic resonance. Via numerical simulations and analytically, 
we study how the signal to noise ratio (SNR) and  power ampliﬁcation ($\eta$) behave
as a function of the model parameters. $\eta$ as well as SNR depicts a pronounced peak 
at particular noise strength $T$. The magnitude of $\eta$ is higher for temperature 
dependent $\gamma$ case. In the presence of $N$ particle,  $\eta$ is considerably 
amplified as $N$ steps up showing the the weak periodic signal plays a vital   role
in controlling the noise induced dynamics of excitable system

In conclusion, in this work, we  explore the crossing rate and stochastic resonance 
of a single as well as many Brownian particles that move in a piecewise linear bistable
potential by considering both temperature dependent and independent viscous friction 
cases.  Although a generic model system is considered,  the present study helps to 
understand the dynamics of excitable systems such neural and cardiovascular systems.

 {\it Acknowledgment.\textemdash} 
 We would like to thank Mulugeta Bekele for the interesting discussions we had. MA 
 would like to thank Mulu Zebene for the constant 
encouragement.




\begin{thebibliography}{60}
\bibitem{am1}  H.A. Kramer. Physica {\bf 7}, 284 (1940).
\bibitem{am2}  P. H¨anggi, P. Talkner and M. Borkovec, Rev. Mod. Phys. {\bf 62}, 251 (1990).
\bibitem{am3}  P.J. Park and W. Sung, J. Chem. Phys. {\bf 111}, 5259 (1999).
\bibitem{am4}  S. Lee and W. Sung, Phys. Rev. E {\bf 63}, 021115 (2001).
\bibitem{am5}  P. H¨anggi, F. Marchesoni and P. Sodano, Phys. Rev. Lett. {\bf 60}, 2563 (1988).
\bibitem{am6}  F. Marchesoni, C. Cattuto and G. Costantini, Phys. Rev. B, {\bf 57}, 7930 (1998).
\bibitem{am7}  P. H¨anggi and F. Marchesoni, Rev. Mod. Phys. {\bf 81}, 387 (2009).
\bibitem{am8}  K.L. Sebastian and Alok K.R. Paul, Phys. Rev. E {\bf 62}, 927 (2000).
\bibitem{am9}  M. Bekele, G. Ananthakrishna, N. Kumar - Physica A  {\bf 270}, 149 (1999).
\bibitem{am10} O. Reynolds,  Phil Trans Royal Soc London { \bf 177}, 157 (1886).
\bibitem{am111} B. Lindnera, J. Garcia-Ojalvob, A. Neimand, L. Schimansky-Geier, Phys. Reports  {\bf 392}, 321 (2004).
\bibitem{am11} W. Chen, M. Asfaw , Y. Shiferaw,  Biophys J {\bf 102}, 461 (2012).
\bibitem{am12} M. Asfaw, E. A. Lacalle, Y. Shiferaw, Plos One, {\bf 8}, e62967 (2013).
\bibitem{am13} R. Benzi, G. Parisi, A. Sutera and A. Vulpiani, Tellus {\bf 34}, 10 (1982).
\bibitem{am14} L. Gammaitoni, P. H¨anggi, P. Jung and F. Marchesoni, Rev. Mod. Phys. {\bf 70}, 223 (1998).
\bibitem{am15} A. Neiman and W. Sung, Phys. Lett. A {\bf 223}, 341 (1996).
\bibitem{am16} P. Jung, U. Behn, E. Pantazelou, and F. Moss, Phys. Rev. A {\bf 46}, R1709 (1992).
\bibitem{am17} J. F. Lindner, B. K. Meadows, W. L. Ditto, M. E. Inchiosa, and A. R. Bulsara, Phys. Rev. Lett. {\bf 75}, 3 (1995); Phys. Rev. E 53, 2081 (1996).
\bibitem{am18} F. Marchesoni, L. Gammaitoni, and A. R. Bulsara, Phys. Rev. Lett. {\bf76}, 2609 (1996).
\bibitem{am19} I. E. Dikshtein, D. V. Kuznetsov, and L. Schimansky-Geier, Phys. Rev. E. {\bf 65}, 061101 (1996).
\bibitem{am20} I. Goychuk and P. Hanggi, Phys. Rev. Lett. {\bf 91}, 070601 (2003).
\bibitem{am21} H. Yasuda et al., Phys. Rev. Lett. {\bf 100}, 118103 (2008).
\bibitem{am22} J. M. G. Vilar and J. M. Rubi, Phys. Rev. Lett. {\bf 78}, 2886 (1997).
\bibitem{am23} J. F. Lindner, M. Bennett, and K. Wiesenfeld, Phys. Rev. E {\bf 73}, 031107
\bibitem{am24} M. Asfaw and W. Sung, EPL {\bf 90}, 3008 (2010).
\bibitem{am25} M. Asfaw, Phys. Rev. E {\bf 82}, 021111 (2010).
\bibitem{am26} C. W. Gardiner. Handbook of Stochastic Methods for Physics, Chemistry and the
Natural Sciences. Springer, Berlin, (1984).

\end{thebibliography}
\end{document}